\documentclass[11pt]{article}

\usepackage{amsmath}
\usepackage{amssymb}
\usepackage{float}
\usepackage{graphicx}

\usepackage{color} 
\usepackage{booktabs}
\usepackage{lineno}
\usepackage{url}
\usepackage{pdfpages}
\usepackage{setspace} 
\usepackage{lineno}
\usepackage{comment}
\usepackage[numbers,square,comma,sort&compress]{natbib}

\usepackage{color} 
\usepackage{tabularx}
\usepackage{lineno}
\usepackage{url}

\usepackage[table,x11names]{xcolor}
\usepackage{blkarray}
\usepackage{amsmath,amstext,amsthm,amssymb}
\usepackage{enumerate}
\usepackage{hyperref}
\usepackage{verbatim}

\usepackage[table]{xcolor}

\usepackage{subcaption}  
\hypersetup{
	bookmarks=true,         
	unicode=false,          
	pdftoolbar=true,        
	pdfmenubar=true,        
	pdffitwindow=false,     
	pdfstartview={FitH},    
	pdftitle={My title},    
	pdfauthor={Author},     
	pdfsubject={Subject},   
	pdfcreator={Creator},   
	pdfproducer={Producer}, 
	pdfkeywords={keywords}, 
	pdfnewwindow=true,      
	colorlinks=true,       
	linkcolor=blue,          
	citecolor=blue,        
	filecolor=magenta,      
	urlcolor=cyan           
}

\newcommand{\commentout}[1]{}

\usepackage[labelfont=bf,labelsep=period,justification=raggedright]{caption}

\makeatletter
\renewcommand{\@biblabel}[1]{\quad#1.}
\makeatother

\date{}

\title{\bf{Social physics in the age of artificial intelligence
} }

 \author{The Anh Han$^{1,\star}$, Joel Z. Leibo$^{2}$, Tom Lenaerts$^{3,4,5}$, Iyad Rahwan$^{6}$, Fernando Santos$^{7}$, \\  Matja{\v z} Perc$^{8,9,10,11,\ddagger}$, Valerio Capraro$^{12,\ddagger,\star}$}

\begin{document}
	\maketitle
	{\footnotesize
		\noindent
		$^{1}$ School of Computing, Engineering and Digital Technologies, Teesside University, Middlesbrough, UK.\\
 $^{2}$ Google DeepMind, London, UK \\
    $^{3} $ Machine Learning Group, Université Libre de Bruxelles, Brussels, Belgium \\
    $^{4}$ AI Lab, Vrije Universiteit Brussel, Brussels, Belgium \\
    $^{5}$ Center for Human-Compatible AI, UC Berkeley, USA \\
  $^{6}$ Max Planck Institute for Human Development, Center for Humans \& Machines, Berlin, Germany\\
  $^{7}$ Informatics Institute, University of Amsterdam, Amsterdam, The Netherlands\\ 
  $^8$ Faculty of Natural Sciences and Mathematics, University of Maribor, Maribor, Slovenia\\
  $^9$ Community Healthcare Center Dr. Adolf Drolc Maribor, Maribor, Slovenia\\
  $^{10}$ Department of Physics, Kyung Hee University, Seoul, Republic of Korea\\
  $^{11}$ University College, Korea University, Seoul, Republic of Korea\\ 
   $^{12}$ University of Milan Biccoca, Milan, Italy\\ \\
  $\ddagger$:  Equal last author 
  
\noindent		$^\star$ Corresponding: The Anh Han (t.han@tees.ac.uk), Valerio Capraro  (valerio.capraro@unimib.it) 
	}

\newpage
\section*{Abstract}
Artificial intelligence (AI) systems are rapidly becoming more capable, autonomous, and deeply embedded in social life. As humans increasingly interact, cooperate, and compete with AI, we move from purely human societies to hybrid human–AI societies whose collective dynamics cannot be captured by existing behavioural models alone. Drawing on evolutionary game theory, cultural evolution, and Large Language Models (LLMs) powered simulations, we argue that these developments open a new research agenda for social physics centred on the co-evolution of humans and machines. We outline six key research directions. First, modelling the evolutionary dynamics of social behaviours (e.g. cooperation, fairness, trust) in hybrid human–AI populations. Second, understanding machine culture: how AI systems generate, mediate, and select cultural traits. Third, analysing the co-evolution of language and behaviour when LLMs frame and participate in decisions. Fourth, studying the evolution of AI delegation: how responsibilities and control are negotiated between humans and machines. Fifth, formalising and comparing the distinct epistemic pipelines that generate human and AI behaviour. Sixth, modelling the co-evolution of AI development and regulation in a strategic ecosystem of firms, users, and institutions. Together, these directions define a programme for using social physics to anticipate and steer the societal impact of advanced AI. \\

\noindent \textbf{Keywords:} Social physics, artificial intelligence, evolutionary game theory, human-AI
interactions, language, delegation, machine culture, prosocial behaviour, AI regulation. 

\newpage
\tableofcontents


\maketitle
\newpage

\section{Introduction}

Artificial Intelligence (AI) systems are advancing rapidly and are becoming increasingly autonomous and embedded in everyday life \citep{andras2018trusting,brinkmann2023machine,rahwan2019machine}. AI promises to transform how we address a wide range of societal challenges, from curing diseases and mitigating climate change to increasing productivity across multiple sectors \citep{ maslej2025artificial}. Yet these benefits come with complex, large-scale risks and challenges, spanning economic, social, and existential concerns \citep{bengio2024managing,mclean2023risks,capraro2024impact,hammond2025multi,bengio2026internationalAISafetyReport}.

In contemporary society, and likely even more so in the near future, humans and AI systems with diverse roles and capabilities coexist and interact, giving rise to co-evolutionary dynamics that differ markedly from those observed in purely human settings \citep{fernandez2022delegation,kobis2025delegation,pedreschi2025human,bonnefon2024moral,hidalgo2021humans}. These interactions can generate system-level effects that are difficult to anticipate, model, and govern, and heterogeneity in AI capabilities and behaviours further complicates prediction and the design of effective interventions \citep{tomasev2025virtual,hadfield2025economy,bengio2026internationalAISafetyReport}.

Existing behavioural research, both theoretical and empirical, has predominantly focused on human-to-human interactions, largely overlooking the distinctive dynamics that arise when AI systems are integrated into hybrid human–AI societies \citep{dafoe2021cooperative,zimmaro2024emergence,brinkmann2023machine,santos2024prosocial}. Evolutionary game theory \citep{hofbauer1998evolutionary}, cultural evolution \citep{mesoudi2016cultural}, and generative AI based simulations \citep{lu2024llms} provide tractable frameworks for analysing strategic adaptation, feedback processes, and population-level outcomes in such settings. In this article, we argue that recent and forthcoming advances in AI define a new set of research questions that are of primary importance for social physics.

We propose six research directions we believe are particularly important to pursue to understand the impact of AI in the evolution of social behaviour (Table \ref{table:summary}). 
First, given the emergence and proliferation of hybrid human-AI societies 
\citep{santos2024prosocial,zimmaro2024emergence,tomasev2025virtual, hadfield2025economy,PERRET2026118130} it is now crucial to revisit established domains of social behaviour (e.g. cooperation, fairness, and trust) \citep{perc2017statistical,nowak2006five,kumar2020evolution} and fundamental political questions of ``how can we live together?'' \citep{dafoe2020open,gabriel2020artificial,leibo2025societal}. 
Second, as intelligent machines become integral to our lives, the study of cultural evolution must be updated to reflect AI's influence on social behaviour  and emergent social norms \citep{mesoudi2016cultural,brinkmann2023machine}.
Third, the recent advances in LLMs and generative AI suggest re-examining how these technologies influence the evolution of language and behaviour \citep{capraro2024language,battiston2025higher}. 
Fourth, as humans delegate more tasks and responsibilities to AI systems, there is a need to explore the evolution of such delegation behaviours, understanding how humans and machines negotiate responsibilities and roles in decision-making processes \citep{deMelo2019human,March2021lessons,fernandez2022delegation,gabriel2024ethics,terrucha2025humans,kobis2025delegation}. 
Fifth, because humans and AI systems rely on fundamentally different epistemic pipelines \citep{loru2025simulation,quattrociocchi2025epistemological,perc2025counterfeit}, we must investigate how selection acts not only on observable behaviour but also on the underlying cognitive and epistemic processes that generate it in hybrid societies. 
Finally, in order to effectively manage  AI risks amid rapid progress \citep{bengio2024managing}, 
 it is crucial to understand the evolutionary dynamics of safe AI development and and how regulation can be strategically applied to drive these dynamics toward beneficial outcomes for society \citep{han2020regulate,han2019modelling}. 

{\tiny
\setlength{\arrayrulewidth}{0.1mm}
\setlength{\tabcolsep}{8pt}
\renewcommand{\arraystretch}{2}
{\tiny
\begin{table}
\begin{center}
\caption{Summary of research directions in social physics in the age of AI.}
\label{table:summary}
  \centering
{\rowcolors{2}{green!80!yellow!20}{green!70!yellow!30}
\begin{tabular}{ |p{2.2cm}|p{7.4cm}|p{6.8cm}|  }
\hline
\textbf{Research Area} & \textbf{Specific Question} & \textbf{Potential Design} \\
\hline
Evolutionary dynamics of social behaviours in hybrid human-AI systems
 & (1) How can we model humans' expectations about AI behaviour in heterogeneous, legal, normative, and hierarchical hybrid systems? (2) How can we connect short-term behavioural experiments with long-term evolutionary dynamics in hybrid populations?   & Evolutionary game  and statistical-physics models (e.g. replicator dynamics, stochastic processes) calibrated with  experiment and field data; hybrid human–AI experiments varying AI fraction, transparency, and coordination between AI agents.  \\
Machine culture: machine-generated and mediated cultural dynamics & (1) How does algorithmic curation of content and social ties affect cultural diversity, norm formation, and exploration–exploitation trade-offs? (2) How do human preferences and platform incentives shape the evolution of machine behaviour and AI-mediated culture?   & Agent-based and network models of cultural transmission under different recommender objectives; A/B tests on engagement vs diversity; simulations with generative models as cultural innovators; analysis of human-in-the-loop training (e.g. RLHF) as cultural selection on AI. \\
Co-evolution of language and behaviour    &(1) How does linguistic framing,  when mediated by LLMs, shape human and AI decisions in strategic interactions? (2) How do language and social behaviour co-evolve in hybrid populations?  &  Language-based game-theoretic models; large-scale framing experiments; societies of LLM powered simulations; meta-analyses linking linguistic features to behaviour.\\
Evolutionary dynamics  of AI delegation  & (1) Does delegation to AI agents increase or decrease propensity towards cooperation, honesty, etc.? (2) How do incentives and AI agent design impact long-term dynamics of cooperation, inequality, honesty norms, etc.? & Simulations using simple versus LLM-based agents; Lab-based and online experiments involving humans and AI agents in dynamic, repeated interaction. \\
Evolution of epistemic pipelines in hybrid human--AI societies   & (1) Under what conditions does selection favour agents with different epistemic pipelines (e.g. grounded vs text-only, causal vs correlational, cautious vs overconfident)? (2) When do hybrid systems converge to epistemic vigilance vs equilibria dominated by persuasive but ungrounded outputs? & Extended evolutionary models where types encode cognitive/epistemic parameters (grounding, counterfactual reasoning, confidence, abstention); simulations of hybrid human–AI populations; experiments varying incentives for verification vs persuasion and measuring downstream trust and misinformation. \\
Co-evolutionary dynamics of AI development and regulation &(1) How do the strategic interactions between firms, users, and regulators shape the trajectory of AI capabilities, safety investments, and power concentration? (2) Which regulatory and institutional interventions can steer this ecosystem toward safe and socially beneficial outcomes?  
&  
Game theory and population dynamics models of AI labs, regulators, and users (arms races vs cooperation; open-source vs proprietary); scenario analysis of policy levers (standards, audits, windfall-sharing);  experiments and LLM-based simulations of AI actors’ decisions.   \\
\hline
\end{tabular}}
\end{center}
\end{table}
}
}

\section{The rise of Artificial Intelligence}
AI technologies vary widely in their complexity and capabilities, spanning a spectrum from traditional rule-based systems to sophisticated models with agency and adaptive learning \citep{bengio2024managing,perrault2024artificial,maslej2023ai}.

At the simplest level are the traditional rule-based AI systems, which perform specific tasks by following predefined rules and scripts (for example, some legacy customer-service bots or menu-based virtual assistants). 
While limited in capabilities, they provide efficient and cost-effective alternatives for promoting prosocial interactions \citep{mu2024multi,santos2024prosocial}.  
A more advanced class of systems comprises chatbots powered by large language models (LLMs), deep neural networks, which can dissect, extrapolate from and generate human-like text \citep{chang2024survey}. Different from the traditional rule-based systems, LLM-based chatbots are not  confined to fixed responses; instead, they can engage in dynamic conversations, offer insights, capable of emulating some cognitive tasks, and can perform a wide range of linguistic tasks. This capability makes them invaluable tools for applications such as customer service, content creation, education, and beyond. They can support humans in interactions with each other and communicate as well with other AI systems and advanced tools.  In the context of game theory and population dynamics, these systems allow formulate language-based games (more discussion on co-evolution of language and behaviour in Section \ref{section:coevolution-language-behaviour}). 
These AI models can learn from interactions, becoming more effective over time by adjusting their responses and behaviours based on new observations..

Given these capabilities, we envision several dimensions along which the rise of AI can impact evolutionary dynamics in human-AI societies.


First, unlike humans, who typically learn from personal experience and culturally transmitted norms, AI systems can process vast datasets rapidly and modify their behaviour at an unprecedented rate \citep{griffiths2020understanding}. When such systems are embedded in social environments, they participate in reciprocal interactions, reputation building, and norm formation alongside humans, thereby potentially altering the mechanisms that sustain cooperation, fairness, and other social behaviours. Even small fractions of artificial agents can  have disproportionate effects on population-level outcomes \citep{terrucha2024art}. These considerations motivate a first research direction, on the evolutionary dynamics of social behaviours in hybrid human–AI systems, where heterogeneous human and AI agents co-adapt through feedback loops in behaviour, expectations, and institutional constraints (Section \ref{section:evolutionar-hybrid}).

Second, AI systems increasingly participate in the generation, transmission, and selection of cultural content \citep{brinkmann2023machine}. Generative models produce novel artefacts and ideas, recommender systems curate and rewire social networks, and LLM-based assistants mediate knowledge transmission at scale. Together, these capabilities turn AI into both a generator and a gatekeeper of culture, with profound implications for cultural diversity, norm formation, and collective exploration–exploitation trade-offs. These issues are central to a second research direction on machine-generated and machine-mediated cultural dynamics (Section \ref{section:machine-culture}).

Third, because many of the most widely deployed AI systems are language-based, they reshape how decisions are framed, justified, and coordinated. LLMs both respond to and produce linguistic descriptions of decision problems, thereby influencing how humans perceive strategic situations and which behaviours they regard as acceptable or moral. This tight coupling between language and choice, mediated by AI, motivates a focus on the co-evolution of language and behaviour in hybrid populations (Section \ref{section:coevolution-language-behaviour}).

Fourth, as AI systems become more capable and useful, humans may increasingly rely on them, facing decisions about what to delegate, to which system, and under which conditions \citep{fernandez2022delegation, kobis2025delegation,terrucha2025humans}. Differences in autonomy, reliability, and transparency across AI systems create heterogeneous delegation patterns, feedback on trust and reliance, and new forms of strategic interaction between humans and AI assistants. Understanding how such delegation decisions emerge and evolve is the focus of a fourth research direction, investigating  the evolutionary dynamics of AI delegation  (Section \ref{section:delegation}).

Fifth, AI systems may make judgments through potentially different epistemic pipelines compared to humans—from grounding and experience to motivation, causal reasoning, and error-monitoring. Behavioural similarity with humans does not necessarily entail cognitive or epistemic equivalence \citep{loru2025simulation,quattrociocchi2025epistemological}. This calls for models in which selection acts not only on observed behaviour but also on underlying epistemic processes, motivating a fifth research direction on the evolution of epistemic pipelines in hybrid human–AI societies (Section \ref{section:epistemic-pipelines}).

Finally, AI systems are developed and deployed within a strategic ecosystem of firms, governments, users, and other stakeholders, whose incentives can generate races, power concentration, or cooperation around safety and standards. Because AI can be scaled and modified rapidly, small changes in incentives or regulation may have large, path-dependent effects on this ecosystem. This motivates a final research direction on the co-evolution of AI development and regulation, using evolutionary and game-theoretic models to identify interventions that steer AI progress towards socially beneficial and safe outcomes (Section \ref{section:Regulation}).


\section{Evolutionary dynamics of social behaviours in hybrid human-AI systems}
\label{section:evolutionar-hybrid}
Human-AI co-adaptation resembles evolutionary dynamics in biological systems. AI applications are trained on human-generated data, which in turn affects human social behaviours and shapes future datasets. Such co-evolving dynamics are evident in multiple applications: in recommender systems, users’ future interests are shaped by current algorithmic recommendations \citep{piao2023human}; future human behaviours can be influenced by socially interactive agents, themselves adapting to match users’ personalities and preferences \citep{lugrin2021introduction}; in the context of generative AI, incentives for humans to put effort in generating high-quality data can change over time, affecting the long-term value of data used to train future models \citep{huang2023generative}; when being classified by machine learning algorithms, humans can change their features, possibly leading to data shifts and motivating algorithms' re-training \citep{hardt2016strategic,couto2025collective}.  The feedback loops characterizing human-AI systems result in social dynamics challenging to describe, anticipate, and control \citep{pedreschi2025human}. Tools used in evolutionary biology and statistical physics can play a major role to overcome this challenge, and inform the design of AI that improves human social behaviours.

Human decision-making processes and interaction contexts can be highly complex. Likewise, as described in Section 2, artificial agents powered by large models are increasingly sophisticated. There is however experimental evidence that incredibly simple agents suffice to impact human behaviour towards cooperative and coordinated outcomes. It was shown that a small fraction of bots with a random behaviour can significantly improve human coordination on networks \citep{shirado2017locally}. In repeated cooperation dilemmas, it was also shown that resilient cooperators (i.e., with a fixed behaviour) can stabilise cooperation even among more self-interested humans who might otherwise defect \citep{mao2017resilient}. 

Conveniently, the evolutionary dynamics introduced by simple agents can be studied through evolutionary dynamics and statistical physics methods, applying tools such as the replicator equation or stochastic process analysis \citep{santos2024prosocial,mu2024multi}. Often, these highly stylized models abstract away the complexity of interactions and focus instead of capturing mathematically the macroscopic impact of introducing simple hard-coded agents in a population of otherwise adaptive individuals.  
Although far from the complexity of current autonomous agents, considering hard-coded agents in a population of otherwise adaptive individuals provides an intuition for the profound impacts of AI agents in the social behaviour of humans. In this regard, existing models reveal that a small fraction of unconditionally cooperative agents suffice to stabilize cooperation \citep{sharma2023small}. Similar models also reveal that the advantages of such unconditional cooperators are contingent on adaptation properties such as intensity of selection \citep{zimmaro2024emergence}. When reputations are available, conditional agents steering cooperation towards well-reputed individuals can trigger large-scale cooperation \citep{pires2025artificial}. This modelling approach has also been applied to study social dynamics in  dilemmas of coordination \citep{guo2023facilitating}, fairness \citep{santos2019evolution,song2026evolution} and collective risk \citep{terrucha2024art,santos2020picky}. 

The previous works provide prolific tools to study evolutionary dynamics of social behaviour in rather simple systems. Applying such methods already informs us that small fractions of artificial agents, even with a fixed strategy or simple behavioural rules, can have a substantial effect on human social behaviour. Nonetheless, real-world hybrid systems reveal features challenging to capture in current models. First, real-world systems are large and highly heterogeneous: humans across different societies vary in how they perceive artificial and adaptive agents \citep{mckee2023humans, bergman2024stela}. Likewise, artificial agents can themselves be highly heterogeneous and vary in shape, communication modality or cognitive abilities. How to capture such heterogeneity in a model? Second, while humans and AI co-evolve, so do legal and normative systems regulating AI. A key example is transparency requirements. AI applications can be required to become increasingly interpretable, and differences in how interpretability is implemented can greatly affect humans’ expectations regarding artificial behaviour and how they adapt to it. How to incorporate, in a model, the evolving legal and normative systems (i.e., system of laws, rules, principles, or standards) where human-AI interactions take place? Third, previous models typically assume that artificial agents are independent. Artificial agents might however reveal coordinated behaviours, being owned by specific companies or acting on behalf of interest groups with competing interests (see section on \textit{Co-evolution of AI development and regulation}). This suggests hierarchical models, where interactions can occur at different levels: between individual agents or between groups and organizations. Finally, models of social behaviour evolution in hybrid systems should be informed — and inform — behavioural experiments, data collection and A/B testing \citep{neto2025cooperation,santos2020picky}. 

To advance our understanding of social behaviour dynamics in hybrid human-AI systems, we believe it is fundamental to overcome the challenge of modelling agents’ heterogeneity, interactions' complexity, evolving AI regulation, dynamics between AI owners, and improving the methods to link theoretical modelling and real-world data. Meeting these challenges boils down to tacking two key questions: 1) How to model humans’ expectations about AI behaviour in heterogeneous, legal, normative, and hierarchical hybrid systems? 2) How to connect short-term behavioural experiments with long-term evolutionary dynamics modelling of hybrid systems?

\section{Machine culture: machine-generated and mediated cultural dynamics}
\label{section:machine-culture}
In this section, we explore the influence of AI on social behaviour through the lens of Culture Evolution \citep{mesoudi2016cultural}. Historically, cultural evolution has been studied through frameworks that emphasize the roles of human cognition, social learning, and communication. With the advent of intelligent machines, these frameworks must be re-examined to accommodate the impact that AI is having on the generation, transmission, and selection of cultural traits.

One of the most significant contributions of AI to cultural dynamics is its capacity to generate novel cultural artifacts and ideas. Traditional models of cultural variation rely on human creativity and the recombination of existing cultural elements. Yet, generative AI systems have begun to operate as independent sources of innovation. As an example of combinatorial novelty, text-to-image generative algorithms, such as DALL-E \citep{johnson2021openai} and Stable Diffusion \citep{rombach2022high} enable artists and designers to rapidly explore novel combinations of visual concepts at an unprecedented scale--e.g. generating hundreds of visual concepts for an avocado-shaped chair in a matter of seconds. Moreover, reinforcement learning systems, exemplified by AlphaGo, have demonstrated the capacity to develop strategies that are not only superior to human-derived ones but also ``alien'' to human culture. The unexpected and highly unconventional move 37 by AlphaGo in its match against Lee Sedol  exemplifies this phenomenon \citep{leibo2019autocurricula}.

Beyond generating new cultural artifacts, AI systems are also fundamentally altering the mechanisms of cultural transmission. Traditionally, cultural information has been passed from one individual to another through direct social interactions, observation, and teaching. However, AI has introduced new pathways for the dissemination of cultural knowledge, first through recommender systems, and more recently through LLMs. AI-driven recommender systems are reshaping the social networks that underlie cultural transmission. These systems, which suggest connections, content, and opportunities based on user behaviour \citep{li2017survey}, are subtly but significantly rewiring the way individuals interact with one another and with cultural artifacts. By prioritizing certain types of content and connections, recommender systems can influence social norms, amplify certain cultural trends, and even create new forms of social capital. This raises critical questions about the long-term implications of such rewiring: How does the algorithmic curation of social networks affect the diversity of cultural expression and the dynamics of cultural transmission? And what are the tradeoffs that platform operators face between user engagement and diversity of cultural transmission \citep{baumann2024optimal}? LLMs, trained on vast corpora of human text, now act as intermediaries in the transmission of knowledge, serving as both reservoirs and conduits of cultural information (e.g. as teachers \citep{lucas2020value}). They facilitate the spread of ideas across individuals and generations, often in ways that are more efficient and far-reaching than traditional means. However, this also means that the cultural content transmitted is increasingly shaped by the biases and structures inherent in these AI systems. As a result, AI is not just a passive transmitter of culture but an active participant that influences what information is emphasized, what is marginalized, and how cultural narratives are constructed.

The selection process in cultural evolution---the mechanism by which certain cultural traits become more prevalent while others fade away---will also be deeply impacted by AI. In the digital age, the algorithms that curate content on platforms like social media, streaming services, and e-commerce sites have become powerful cultural gatekeepers. These algorithms are designed to maximize user engagement, often by selecting content that aligns with users' past behaviours and preferences. However, in doing so, they play a significant role in shaping cultural landscapes, determining which ideas gain traction and which are relegated to obscurity \citep{diresta2024invisible}. The implications for cultural evolution may be profound \citep{brinkmann2023machine}. Algorithms, by filtering and prioritizing content, are not just responding to user preferences---they are actively shaping them. This is indeed the explicit goal of some recommendation systems, such as link recommendation algorithms \citep{su2016effect}. These algorithms can thus alter the overall structure of social networks, thus impacting the tradeoffs between collective exploration and exploitation \citep{lazer2007network,mason2012collaborative}. These algorithms may also create feedback loops where certain cultural traits are continuously reinforced, potentially leading to the homogenization of culture or the entrenchment of specific norms and values \citep{kizhner2021digital}. 

Another evolutionary selection process is also taking place, in which humans select among AI algorithms. Thus, human preferences shape the evolution of machine behaviour, e.g. by explicitly training LLMs using human feedback \citep{ouyang2022training}, or by simply favouring certain commercial or open source LLMs over others. Since these LLMs subsequently interact with humans at scale, this process shapes both human and machine cultural representations, but also leads to feedback loops \citep{veselovsky2023artificial}. Understanding these processes is essential, in order to avoid degenerate phenomena like 'model collapse' \citep{shumailov2023curse, duenez2023social}.

\section{The co-evolution of language and behaviour}
\label{section:coevolution-language-behaviour}
In the past two decades, a growing body of experimental research with human participants has demonstrated the impact of linguistic content on decision-making. Essentially, the way in which the decision context is described can significantly alter people’s choices (see \citep{capraro2024outcome}, for a review). As LLM-based chatbots become increasingly embedded in everyday activities, individuals are more likely to interact with these systems directly or to make decisions with their assistance. Because these AI systems operate primarily through language, the options they suggest—and, in some cases, the decisions they themselves produce—depend sensitively on how a decision problem is formulated \citep{veselovsky2025localized}. This feature underscores the growing importance of understanding how linguistic framing shapes human behaviour, machine behaviour, and their interaction. We argue that the integration of LLM-based chatbots into human decision-making processes will reshape this research area in at least two ways. First, it will create a heightened demand for formal, mathematical models capable of capturing the effects of linguistic framing on choice (e.g.~\citep{leibo2024theory}). Second, it will open new empirical and practical avenues for studying how language and behaviour co-evolve within human societies, particularly in contexts where human and artificial agents jointly participate in decision-making.

Regarding the first point, existing research has largely concentrated on experimentally demonstrating the influence of language on human decision-making. For instance, a seminal study by Liberman and colleagues \citep{liberman2004name} found that simply changing the label of a prisoner’s dilemma from “Wall Street game” to “Community Game” significantly increased the rate of cooperation among participants. Eriksson and co-authors \citeyearpar{eriksson2017costly} conducted a series of experiments on the ultimatum game, revealing that the way rejection actions are phrased can influence responders’ behaviour. In recent years, several studies have conceptually replicated and expanded these findings into several domains, including the prisoner’s dilemma \citep{engel2014does,mieth2021moral}, the equity-efficiency trade-off game \citep{capraro2018right,huang2019choosing,huang2020maxims}, the dictator game \citep{capraro2019power,chang2019rhetoric,kuang2024language}, corruption games \citep{scigala2022corrupting}, and market games \citep{jimenez2023conditioning,alger2024doing}. In parallel, a smaller but growing theoretical literature has sought to formalize the role of language in strategic interaction, either through language-based games \citep{bjorndahl2013language,bjorndahl2021language,bjorndahl2023sequential} or through the development of language-based utility functions \citep{capraro2024language}.

Research on the behaviour of LLM-based chatbots is still in its early stages. An emergent line of literature has begun investigating how chatbots make decisions in standard economic games \citep{chen2023emergence,dillion2023can,horton2023large,mei2024turing}. A related line of work studies the behaviour of LLM chatbots in more open-ended strategic situations conveyed by simulations of collaborative storytelling protocols (tabletop role-playing games) \citep{vezhnevets2023generative, smith2025evaluating}. With respect to linguistic framing, existing evidence suggests that LLM-based chatbots often exhibit behavioural patterns that are qualitatively similar to those observed in humans; however, important quantitative differences remain. In particular, the relative frequencies with which different actions are selected can diverge substantially from human behaviour \citep{capraro2025publicly}. Taken together, these findings point to two clear gaps in the literature: a limited understanding of how linguistic framing systematically shapes the behaviour of LLM-based chatbots, and the absence of well-developed theoretical models capable of capturing and explaining these effects.

Concerning the first point, an intriguing direction for future research involves exploring how evolutionary game theory can be leveraged to gain a deeper understanding of how linguistic interactions influence the evolution of social behaviours and, vice versa, how the evolution of social behaviour can influence the evolution of language. Evolutionary game theory examines how strategies evolve over time within populations. It has been applied to study the evolution of cooperative behaviours \citep{perc2017statistical,han2025cooperation} and other moral behaviours, such as honesty \citep{capraro2019evolution} and trust \citep{kumar2020evolution,han2021or,PERRET2026118130}. Separately, it has also investigated the evolution of language \citep{puglisi2008cultural,baronchelli2006sharp,danovski2022evolutionary}. However, it is also plausible that language and behaviour co-evolve. Consider a scenario where two agents engage in a verbal interaction before making a decision. During this conversation, they can influence each other’s perception of the decision problem. As a result, their final decision is not just a function of the available economic options but also of the preceding verbal interaction \citep{clark2023meaningful,gaffal2024negotiation}. This dynamic can be repeated across multiple rounds, with the same or different pairs of agents, leading to a co-evolution of language and behaviour. Evolutionary game theory could provide valuable insights into how language and behaviour have evolved together in human societies.

Regarding the second gap, LLM-based chatbots themselves may play a crucial role. An emerging line of literature shows that LLM-chatbots can be a useful instrument to measure the sentiment tenor of a piece of text \citep{rathje2024gpt,feuerriegel2025using}. Building on this approach, a recent meta-analysis of 61 dictator game experiments with human participants shows that sentiment scores generated by GPT-4 significantly explain average behavioural patterns. Specifically, behaviour can be captured by an index that mathematically combines the sentiment associated with three pivotal actions: giving nothing, splitting the endowment equally, and giving all \citep{capraro2024language}. This line of research therefore illustrates how LLM-chatbots themselves can help formally incorporating linguistic descriptions of decision contexts into utility functions, thereby providing a tractable mathematical framework for linking language to behaviour.

While the study of the co-evolution of language and behaviour is inherently fascinating and largely independent of AI, we propose that AI, especially LLM-based chatbots, can play a crucial role in investigating this issue. To illustrate, we conclude this section offering one concrete example. Imagine a situation in which a population of chatbots interacts using verbal communication before making decisions, simulating a potential co-evolution of language and behaviour. Since chatbots evolve much faster than humans, using them to simulate societies could allow us to quickly gain insights into the co-evolutionary processes that took millennia for humans to develop. For instance, in positive-sum games, language might stabilize as an effective means for individuals to signal cooperative intent. This could explain why moral narratives across cultures often emphasize cooperative behaviours \citep{curry2019good}. In contrast, during zero-sum interactions, language might be less beneficial and could even be seen as manipulative, with players interpreting the same words differently, possibly due to conflicting interests.
\section{Evolutionary dynamics of AI delegation}
\label{section:delegation}
An aspect fundamental to the organisation and success of a society is the mechanism of delegating control or execution of tasks to those that are more skilled or more knowledgeable, allowing members to achieve goals beyond their personal capabilities.  This idea of a principal delegating authority to an agent has been studied for many decades \citep{jensen1976theory,shapiro2005agency, gailmard2014accountability} and within a variety of contexts, focussing mostly on answering questions on how to ensure alignment between a principal and an agent when goals may diverge and information and risks may be distributed unequally. With the advent of highly performing AI systems, this line of research has gradually been expanding to the delegation of agency and decision-making authority to these systems \citep{moloi2020agency}, especially since they now have the capacity to outperform humans in different tasks, even in an autonomous setting.  This advancement introduces a new level of complexity in the  dynamics of our societal organisation, requiring one to understand when and how humans should delegate to AI, whether beneficial or detrimental outcomes are more likely, and what the long-term social and socio-technical effects of this delegation may be.  

Different experimental as well as theoretical contributions have been made, and the interest in this topic has been rapidly increasing \citep{lubars2019ask,Canadrian2022rise,hemmer2023human,yankouskaya2026lets}. A very recent theoretical example was provided in the context of participatory budgeting \citep{Shah2025votingdelegate}, which focussed on the question of whether artificial delegates that learn the voting behaviour of participants could be used to replace absent voters to ensure correct representation in perpetual voting systems \citep{lackner2020perpetual,lackner2023proportional}.  The authors demonstrated under a variety of model parameters that artificial delegation ensures representation across some key dimensions;  influence is more equally
distributed, minority voices are satisfied more regularly, and group entitlements are better respected. Crucially, the introduction of artificial delegates produces results that align with those under full turnout, thus ensuring artificial delegates rarely lead to outcomes which would not have won in their absence. 

This beneficial role of delegation to autonomous, while constrained, agents has been demonstrated also in depth in a series of behavioural experiments. For example, \cite{demelo2018social} showed how people act more fairly in bargaining games when acting through a programmed agent. The same positive effect on the level of cooperation was revealed within the context of a series of framed public goods game experiments. Similarly, \cite{fernandez2022delegation} reported, within the context of a collective risk dilemma (CRD), where participants need to achieve a goal at the risk of incurring a loss if the goal is not achieved, that either selecting a pre-defined agent or programming the agent's decision-making behaviour augments group success significantly.  These works demonstrate that delegation to autonomous agents seems to work as a form of behavioural commitment \citep{brocas2004commitment}, which prevents participants to deviate to another course of actions while the game is unfolding and thus minimising the effect of emotional responses based on the observed behaviour of others \citep{frank88}. Even when revision of the programmed strategy is possible, the participants maintain the ambition to reach the target and appear to be less influenced by a negative outcome in the first round of the game compared to a humans-only experiment \citep{terrucha2025humans}. An evolutionary game theory model differentiating between human and delegated strategy dynamics, based on when and how decision/programming errors can be made, provides a tentative explanation for the improved success in the CRD \citep{terrucha2024committing}. Interestingly, the balance between a triad of fair, compensating and reciprocating strategies \citep{fernandez2022delegation, domingos2020timing, domingos2021modeling,terrucha2025humans} in the CRD appears to determine when delegation is more successful or when delegation is more likely to be adopted. In a hybrid scenario simulated in this model where both delegation and no-delegation are possible, delegation, specifically the selection of a pre-defined agent, emerges as most beneficial in the long run, confirming observations made in  \citep{Canadrian2022rise}.  

Notwithstanding these promising and positive results, several issues undermine the use of AI systems as autonomous delegates.  In the context of the aforementioned CRD work, significant inequality in the individual gains that each participant receives at the end of the experiment were observed, which may explain why the post-experiment questionnaire revealed that most participants would prefer to play themselves.  Such high degrees of inequality may hinder adoption of AI delegation as a solution, even when they could be most useful to address risky decision-making problems. Additionally, several works, even in the classic principal-agent literature, have revealed that delegation may also lead to more selfish and even cheating behaviour.  \cite{March2021lessons} reviewed more than 160 experimental studies with computer players and while one of his main conclusions was also that participants change their behaviour when interacting with computer players, i.e., they behaved more rationally and selfishly, he also observed that if subjects are aware of the interaction with a computer player, they may also learn to exploit them, which is more problematic.   

This issue of exploitation also recurs in work on human-AI interaction. For example, \cite{crandall2018cooperating} showed that human participants cooperated less with autonomous agents than with other humans, an observation confirmed in \citep{ishowo2019behavioural}. Moreover, participants were keen to exploit their artificial counterparts, who they assumed to be as cooperative as humans, or that it is acceptable to exploit them \citep{karpus2021algorithm}. If rules require AI systems to be cooperative for each individual user, their exploitation may become the rule, undermining the mutual benefit that many people are counting on from human-AI interactions.  This effect was also observed in an evolutionary model wherein an adapting population of agents interacts with a population of pre-set probabilistic agents, which serve as a proxy for AI agents; the individuals in the adapting population reduced their cooperativeness if the autonomous agents were perceived to be sufficiently cooperative \citep{terrucha2024art}. 

In the context of delegation, the risk of dishonest behaviours, both at the principal and agent side, were extensively analysed by \cite{kobis2025delegation}.  The authors showed through a series of experiments that delegation will increase the likelihood of dishonest behaviour, with humans preferring their agents to covertly act in a dishonest manner when it is implicit in the agents' design. Their expansion to LLM agents, which demonstrate a wider behavioural repertoire than simple rule-based systems, further supports this observation. With the increase in using artificial delegates, the amount of dishonest or unethical behaviour will most likely increase, especially since AI systems will tend to comply with such requests due to the way they are currently designed to please the user.  Dishonesty or deception from the side of LLM agents was also found by \cite{hagendorff2024deception}. The authors showed that LLM-based agents may induce false beliefs in other agents, allowing the former to reap a benefit at cost to the other.  Within the context of human-AI delegation, or even AI-AI delegation, such scenarios are clearly problematic, potentially generating distrust in such systems and hindering the deployment of AI agents in cases where they could really be beneficial. Understanding thus the conditions under which deceptive behaviour evolves and can be counteracted will be important for the deployment of AI agents \citep{sarkadi2021evolution}.

Together, these observations call for more in depth research on the evolutionary dynamics of AI delegation, focusing on mechanisms, either implemented at design-time or as a way to control agents’ behaviour at run-time, that i) alleviate the discussed problems that are inherent to human-AI delegation and ii) boost further the benefits of human-AI and AI-AI delegation. Evolutionary models, incorporating for instance insights obtained from the principal-agent framework, will provide understanding and allow one to ask what-if questions to gauge what may work and what will not. These models will need to be validated through experiments, allowing for the verification of model predictions and evaluating the correctness of the underlying assumptions.  In all cases, human non-rational behaviour will need to be integrated in the studies to ensure that conclusions can support real-world deployment. What remains clear is that advanced AI systems will become part of the different societal organisations humans participate in, requiring thus rapid investments to avoid problems both in the short and long term.

\section{Evolution of epistemic pipelines}
\label{section:epistemic-pipelines}
Much of the literature that applies evolutionary game theory to the study of social behaviour models agents as \emph{carriers of strategies}, whose evolution is governed by the relative success of observable outputs. In these models, selection operates on actions and payoffs, typically formalised through replicator dynamics, best-response dynamics, or related population-level update rules \citep{smith1982evolution,weibull1997evolutionary,hofbauer1998evolutionary,sandholm2010population}. This output-centred approach has proven highly successful in explaining the emergence of many social behaviours. However, this tradition largely abstracts away from the cognitive mechanisms that generate behaviour. This abstraction may become problematic in hybrid human–AI populations, where identical outputs may arise from different internal processes \citep{quattrociocchi2025epistemological,loru2025simulation,perc2025counterfeit}. 

Human decisions may be regarded as emerging from a multi-stage epistemic pipeline that integrates grounding, parsing, experience, motivation, causality, metacognition, and value. Humans perceive and act in a world to which they are physically and socially grounded, parse information through meaning-laden representations, accumulate first-person experience with real consequences, reason causally about counterfactuals, act under endogenous motivations and values, and monitor their own uncertainty and epistemic limits. By contrast, the epistemic pipeline of contemporary AI systems may differ from human epistemic pipelines in some important ways. This is particularly evident in LLMs, where outputs are generated from largely ungrounded linguistic representations and, at best, limited forms of multimodal input—most notably vision and, occasionally, audio—while entirely lacking olfaction, proprioception, interoception, vestibular sensing, and other bodily modalities central to human cognition; statistical parsing is based on tokenization, which is blind to speaker intention, emotional tone, and situational nuance, and merely maps character strings onto numerical indices; any prior experience, when present, is still ungrounded, lacks real stakes, and is not value-rich, as it excludes harm, vulnerability, and mortality; inference is primarily driven by statistical correlations, and when LLMs engage in causal inference, their performance typically falls short of human reasoning \citep{chi2024unveiling}, lack of intrinsic motivation or value, and limited metacognitive access to uncertainty \citep{quattrociocchi2025epistemological}. These differences imply that behavioural equivalence does not entail epistemic or cognitive equivalence, even when AI and humans appear to act similarly in social settings \citep{quattrociocchi2025epistemological,loru2025simulation,perc2025counterfeit}.

For this reason, it is increasingly important to develop evolutionary models in which selection acts not only on overt behavioural phenotypes, but also on the underlying cognitive processes that generate them, consistent with the fact that biological evolution operates on heritable mechanisms rather than behaviour per se. One way to do so is to represent agent types by parameters that encode distinct cognitive steps, such as access to grounded feedback versus text-only signals, the capacity for causal or counterfactual reasoning, or the ability to withhold responses under uncertainty. These parameters may themselves evolve or be institutionally shaped, particularly as AI systems are augmented with retrieval mechanisms \citep{lewis2020retrieval}, external tools \citep{schick2023toolformer}, or human feedback \citep{ouyang2022training}; while such augmentations introduce functional analogues
 of human practices, they alter the decision pipeline without necessarily endowing the system with human-like capacity to evaluate and revise beliefs based on evidence.

Previous research has leveraged evolutionary game theory to formalize specific cognitive mechanisms underlying social behaviour. In particular, several studies have examined how intention recognition \citep{han2011intention,han2012corpus,han2015synergy}, theory of mind \citep{stahl1993evolution,devaine_theory_2014,lenaerts2024evolution,wu2025minds,saponara2025evolution}, counterfactual reasoning \citep{pereira2018counterfactual,couto2022introspection,fernandes2024counterfactual}, and intuitive decision making \citep{bear2016intuition,bear2017co,jagau2017general,montero2022fast} shape cooperative behaviour. These contributions demonstrate that evolutionary models can fruitfully incorporate cognitively rich assumptions, moving beyond purely payoff-driven agents. However, this literature remains largely centred on the evolution of cooperation, leaving other socially relevant behaviours unexplored. Moreover, existing models typically assume homogeneous populations and do not account for hybrid societies in which humans interact strategically with artificial agents endowed with distinct cognitive architectures. 

Explicitly formalising cognitive processes in human–AI interactions can generate phenomena that output-based models cannot capture. Consider intuitive versus deliberative moral decision making: even when both processes produce the same immediate moral choice, they need not give rise to the same evolutionary dynamics. Learning, generalization, and justification often depend on the underlying cognitive pathway rather than on behaviour alone, because updating, abstraction, and explanation partly operate over internal representations rather than only on over observed actions. So identical outputs can produce distinct feedback signals and updating processes, leading to divergent trajectories over time. Consequently, models that collapse cognition into outputs risk overlooking key sources of evolutionary divergence in human–AI interactions.

A second example concerns information sharing, credibility, and misinformation. Humans frequently rely on heuristics such as fluency and expressed confidence when evaluating the reliability of information \citep{begg1992dissociation,dechene2010truth,price2004intuitive}. These heuristics are known to be exploitable: some human agents deliberately produce fluent and confident statements despite lacking evidential support, and such strategies can be individually advantageous. LLM-based chatbots can likewise generate fluent and confident outputs even when factually false \citep{kalai2024calibrated,xu2024hallucination}. These mechanisms can be instantiated at scale, with low marginal cost, and weak coupling to verification or reputational consequences. In hybrid populations, this asymmetry can shift selection pressures toward agents—human or artificial—that prioritize persuasive linguistic output over costly verification, favouring equilibria in which linguistic plausibility substitutes for epistemic reliability, a condition recently described as epistemia \citep{quattrociocchi2025epistemological,loru2025simulation}. A question for future research is therefore to understand under which conditions—such as population composition, feedback latency, or incentive structures—hybrid human–AI systems evolve toward epistemic vigilance rather than toward equilibria dominated by persuasive but weakly grounded outputs.

\section{Co-evolution of AI development and regulation}
\label{section:Regulation}
Effective regulation is crucial to ensure responsible and safe development while fostering user trust and adoption \citep{cohen2024regulating,powers2023stuff}. 
Faced with the swift and unpredictable evolution of AI development and  coupled with  accelerated investment in this technology, the urgency to ``deepen our understanding of potential risks and identify actions to address them" \citep{bletchley_declaration_2023} is especially emphasised, as highlighted in the conclusions of the inaugural World Summit on AI. However, despite several proposals on AI regulation aiming to inform strategies, there are little attempts to quantify the potential impacts on near and long-term outcomes \citep{bengio2024managing,hadfield2023regulatory,zhang2021ethics,roberts2023governing,capraro2024impact,dignum2025roadmap}.

Population dynamics approaches, such as evolutionary game theory, have found extensive application in examining decision-making dynamics during various catastrophic challenges like climate change \citep{milinski2008collective,santos2011risk} and nuclear war \citep{baliga2004arms}.
However, the co-evolution of AI development and regulation is fundamentally different \citep{pamlin2015global,han2019modelling}. In disaster analyses related to climate change the central focus is typically on participants’ reluctance to bear personal costs for collectively desired outcomes, reflecting a shared, collective risk borne by all parties \citep{santos2011risk}. By contrast, AI development often exhibits winner-takes-most dynamics, in which successful actors gain substantial relative advantages and face more individualised risk profiles \citep{armstrong2016racing}. The strategic landscape of AI also differs from nuclear arms races: for advanced AI, many of the most severe risks may fall first and most directly on its own developers and immediate users, whereas nuclear powers are generally less exposed to catastrophic harms originating unintentionally from their own arsenals \citep{ord2020precipice,pamlin2015global}. This distinctive configuration of incentives, benefits, and exposure calls for tailored evolutionary and game-theoretic models of AI development and regulation.

There is an emerging body of work that seek to model the strategic dynamics of AI development and impact of regulation \citep{han2021mediating,cimpeanu2022artificial,armstrong2016racing,han2022voluntary,emery2023uncertainty,bova2024both,alalawi2024trust,balabanova2025media,da2025can}. Yet, notable gaps persist that need addressing. 

First, existing models largely focus on winner-takes-all and AI arms race scenarios. However, the trajectory of AI development is not invariably bound by such patterns. Advanced AI may instead support multiple coexisting solutions, e.g. the co-existence of open-source and proprietary foundation models (such as Llama-style and Mistral-style models alongside commercial APIs), which specialise in different domains, pricing structures, and deployment settings rather than converging on a single dominant system \citep{chang2024survey}. Likewise, AI development need not follow arms race dynamics when institutional and market structures favour cooperation over competition—for instance, cross-lab consortia for shared safety benchmarks and evaluation platforms, joint red-teaming efforts, or proposals such as the Windfall Clause, which aims to share extraordinary AI profits more broadly \citep{o2020windfall}. These settings naturally raise key research questions such as:  how do shared safety infrastructures change firms’ incentives to invest in risky capabilities; and when do profit-sharing or benefit-sharing schemes actually reduce competitive pressure in practice?


Second, existing models neglect the social context surrounding AI development, rendering them incomplete. The dynamics and pace of AI progress are likely influenced by the perceived trustworthiness and risks that users attribute to deployed systems—for instance, whether consumers adopt AI decision recommendation in finance or healthcare, or whether organisations are willing to integrate foundation models into critical workflows \citep{andras2018trusting,alalawi2024trust,zhang2021ethics}. As AI becomes more integrated into everyday life, capturing this broader social context—including users’ behaviours, institutional incentives, and societal pressures—becomes essential \citep{brinkmann2023machine,powers2023stuff,leibo2024theory}. For example, user expectations and regulatory scrutiny can influence model design choices and disclosure practices in advanced AI \citep{zhan2023deceptive}. These kinds of settings naturally motivate future research questions, including: under what conditions do users’ trust and adoption decisions slow down or accelerate risky AI races; how do public scandals or highly publicised failures reshape developers’ incentives; and which regulatory or institutional arrangements most effectively align firms’ competitive goals with users’ and society’s safety concerns?

\section{Conclusions}
The emergence of advanced AI marks a turning point in the evolution of our social systems. As hybrid human–AI societies take shape, the collective outcomes we observe will depend not only on technical progress, but on how we design, deploy, and govern these systems. The six research directions we have outlined, spanning  social behaviours, machine culture, language, delegation, epistemic pipelines, and regulation, frame a social-physics research programme for understanding and steering this co-evolution. We have  highlighted key research questions and design approaches for each direction (Table \ref{table:summary}).

 By developing computational models and AI-powered simulations, integrating them with experiments and data, and embedding them in public debate, social physics can help anticipate emergent risks, identify promising opportunities, and inform institutional designs that align AI progress with broadly shared social goals. The sooner we understand the evolutionary forces at work in hybrid human–AI societies, the better placed we will be to steer these systems towards outcomes that are both beneficial and sustainable in the long run.

\section*{Acknowledgement }
T.A.H. is supported by EPSRC (Grant EP/Y00857X/1). T.L. gratefully acknowledges the research support by the F.R.S-FNRS (project grant 40007793), the Service Public de Wallonie Recherche (grant 2010235-ARIAC) by DigitalWallonia4.ai and the Flemish Government through the AI Research Program. M.P. is supported by the Slovenian Research and Innovation Agency (Grant P1-0403).

\bibliographystyle{unsrt}

\bibliography{refs}

@article{ishowo2019behavioural,
  title={Behavioural evidence for a transparency--efficiency tradeoff in human--machine cooperation},
  author={Ishowo-Oloko, Fatimah and Bonnefon, Jean-Fran{\c{c}}ois and Soroye, Zakariyah and Crandall, Jacob and Rahwan, Iyad and Rahwan, Talal},
  journal={Nature Machine Intelligence},
  volume={1},
  number={11},
  pages={517--521},
  year={2019},
  publisher={Nature Publishing Group UK London}
}

@article{kizhner2021digital,
  title={Digital cultural colonialism: measuring bias in aggregated digitized content held in Google Arts and Culture},
  author={Kizhner, Inna and Terras, Melissa and Rumyantsev, Maxim and Khokhlova, Valentina and Demeshkova, Elisaveta and Rudov, Ivan and Afanasieva, Julia},
  journal={Digital Scholarship in the Humanities},
  volume={36},
  number={3},
  pages={607--640},
  year={2021},
  publisher={Oxford University Press}
}

@article{PERRET2026118130,
title = {Disentangling trust from cooperation: Evolution of trust as reduced monitoring in social dilemmas},
journal = {Chaos, Solitons \& Fractals},
volume = {208},
pages = {118130},
year = {2026},
issn = {0960-0779},
doi = {https://doi.org/10.1016/j.chaos.2026.118130},
url = {https://www.sciencedirect.com/science/article/pii/S0960077926002717},
author = {Cedric Perret and The Anh Han and Elias {Fernández Domingos} and Theodor Cimpeanu and Simon T. Powers}
}

@article{chi2024unveiling,
  title={Unveiling causal reasoning in large language models: Reality or mirage?},
  author={Chi, Haoang and Li, He and Yang, Wenjing and Liu, Feng and Lan, Long and Ren, Xiaoguang and Liu, Tongliang and Han, Bo},
  journal={Advances in Neural Information Processing Systems},
  volume={37},
  pages={96640--96670},
  year={2024}
}

@article{crandall2018cooperating,
  title={Cooperating with machines},
  author={Crandall, Jacob W and Oudah, Mayada and Tennom and Ishowo-Oloko, Fatimah and Abdallah, Sherief and Bonnefon, Jean-Fran{\c{c}}ois and Cebrian, Manuel and Shariff, Azim and Goodrich, Michael A and Rahwan, Iyad},
  journal={Nature communications},
  volume={9},
  number={1},
  pages={233},
  year={2018},
  publisher={Nature Publishing Group UK London}
}

@article{brocas2004commitment,
  title={Commitment devices under self-control problems: An overview},
  author={Brocas, Isabelle and Carrillo, Juan D and Dewatripont, Mathias},
  journal={The Psychology of economic decisions},
  volume={2},
  pages={49--67},
  year={2004},
  publisher={Oxford University Press on Demand}
}

@article{shapiro2005agency,
  title={Agency theory},
  author={Shapiro, Susan P},
  journal={Annu. Rev. Sociol.},
  volume={31},
  number={1},
  pages={263--284},
  year={2005},
  publisher={Annual Reviews}
}

@article{gailmard2014accountability,
  title={Accountability and principal-agent theory},
  author={Gailmard, Sean and others},
  journal={The Oxford handbook of public accountability},
  pages={90--105},
  year={2014},
  publisher={Oxford University Press Oxford}
}

@article{jensen1976theory,
  title={Theory of the Firm: Managerial Behavior, Agency Costs and Ownership Structure},
  author={Jensen, Michael C and Meckling, William H},
  journal={Journal of Financial Economics},
  volume={3},
  number={4},
  pages={305--360},
  year={1976}
}

@article{domingos2021modeling,
  title={Modeling behavioral experiments on uncertainty and cooperation with population-based reinforcement learning},
  author={Fern{\'a}ndez Domingos, Elias  and Gruji{\'c}, Jelena and Burguillo, Juan C and Santos, Francisco C and Lenaerts, Tom},
  journal={Simulation Modelling Practice and Theory},
  volume={109},
  pages={102299},
  year={2021},
  publisher={Elsevier}
}

@article{domingos2020timing,
  title={Timing uncertainty in collective risk dilemmas encourages group reciprocation and polarization},
  author={Fern{\'a}ndez Domingos, Elias and Gruji{\'c}, Jelena and Burguillo, Juan C and Kirchsteiger, Georg and Santos, Francisco C and Lenaerts, Tom},
  journal={Iscience},
  volume={23},
  number={12},
  year={2020},
  publisher={Elsevier}
}

@article{sarkadi2021evolution,
  title={The evolution of deception},
  author={Sarkadi, {\c{S}}tefan and Rutherford, Alex and McBurney, Peter and Parsons, Simon and Rahwan, Iyad},
  journal={Royal Society open science},
  volume={8},
  number={9},
  year={2021},
  publisher={The Royal Society}
}

@inproceedings{lackner2020perpetual,
  title={Perpetual voting: Fairness in long-term decision making},
  author={Lackner, Martin},
  booktitle={Proceedings of the AAAI conference on artificial intelligence},
  volume={34},
  number={02},
  pages={2103--2110},
  year={2020}
}

@inproceedings{lackner2023proportional,
  title={Proportional decisions in perpetual voting},
  author={Lackner, Martin and Maly, Jan},
  booktitle={Proceedings of the AAAI Conference on Artificial Intelligence},
  volume={37},
  number={5},
  pages={5722--5729},
  year={2023}
}

@article{lubars2019ask,
  title={Ask not what AI can do, but what AI should do: Towards a framework of task delegability},
  author={Lubars, Brian and Tan, Chenhao},
  journal={Advances in neural information processing systems},
  volume={32},
  year={2019}
}

@article{montero2022fast,
  title={Fast deliberation is related to unconditional behaviour in iterated Prisoners’ Dilemma experiments},
  author={Montero-Porras, Eladio and Lenaerts, Tom and Gallotti, Riccardo and Grujic, Jelena},
  journal={Scientific Reports},
  volume={12},
  number={1},
  pages={20287},
  year={2022},
  publisher={Nature Publishing Group UK London}
}

@article{karpus2021algorithm,
  title={Algorithm exploitation: Humans are keen to exploit benevolent AI},
  author={Karpus, Jurgis and Kr{\"u}ger, Adrian and Verba, Julia Tovar and Bahrami, Bahador and Deroy, Ophelia},
  journal={Iscience},
  volume={24},
  number={6},
  year={2021},
  publisher={Elsevier}
}

@article{hagendorff2024deception,
  title={Deception abilities emerged in large language models},
  author={Hagendorff, Thilo},
  journal={Proceedings of the National Academy of Sciences},
  volume={121},
  number={24},
  pages={e2317967121},
  year={2024},
  publisher={National Academy of Sciences}
}

@article{yankouskaya2026lets,
  title={Who lets AI take over? Cross-national variation in willingness to delegate socially important roles to artificial intelligence},
  author={Yankouskaya, Ala and Almourad, Mohamed Basel and Liebherr, Magnus and Beyahi, Fahad and Xu, Guandong and Ali, Raian},
  journal={AI \& SOCIETY},
  pages={1--19},
  year={2026},
  publisher={Springer}
}

@inproceedings{hemmer2023human,
  title={Human-AI collaboration: the effect of AI delegation on human task performance and task satisfaction},
  author={Hemmer, Patrick and Westphal, Monika and Schemmer, Max and Vetter, Sebastian and V{\"o}ssing, Michael and Satzger, Gerhard},
  booktitle={Proceedings of the 28th international conference on intelligent user interfaces},
  pages={453--463},
  year={2023}
}

@incollection{moloi2020agency,
  title={The agency theory},
  author={Moloi, Tankiso and Marwala, Tshilidzi},
  booktitle={Artificial Intelligence in Economics and Finance Theories},
  pages={95--102},
  year={2020},
  publisher={Springer}
}

@article{Canadrian2022rise,
title = {Rise of the machines: Delegating decisions to autonomous AI},
journal = {Computers in Human Behavior},
volume = {134},
pages = {107308},
year = {2022},
issn = {0747-5632},
doi = {https://doi.org/10.1016/j.chb.2022.107308},
url = {https://www.sciencedirect.com/science/article/pii/S0747563222001303},
author = {Cindy Candrian and Anne Scherer}
}

@article{March2021lessons,
title = {Strategic interactions between humans and artificial intelligence: Lessons from experiments with computer players},
journal = {Journal of Economic Psychology},
volume = {87},
pages = {102426},
year = {2021},
issn = {0167-4870},
doi = {https://doi.org/10.1016/j.joep.2021.102426},
url = {https://www.sciencedirect.com/science/article/pii/S0167487021000593},
author = {Christoph March}
}

@article{deMelo2019human,
  title={Human cooperation when acting through autonomous machines},
  author={de Melo, Celso M and Marsella, Stacy and Gratch, Jonathan},
  journal={Proceedings of the National Academy of Sciences},
  volume={116},
  number={9},
  pages={3482--3487},
  year={2019},
  publisher={National Academy of Sciences}
}

@article{saponara2025evolution,
  title={Evolution favours positively biased reasoning in sequential interactions with high future gains},
  author={Saponara, Marco and Fern{\'a}ndez Domingos, Elias and Pacheco, Jorge M and Lenaerts, Tom},
  journal={Journal of the Royal Society Interface},
  volume={22},
  number={229},
  pages={20250153},
  year={2025},
  publisher={The Royal Society}
}

@inproceedings{Shah2025votingdelegate,
author = {Shah, Apurva and Abels, Axel and Now\'{e}, Ann and Lenaerts, Tom},
title = {Artificial Delegates Resolve Fairness Issues in Perpetual Voting with Partial Turnout},
year = {2025},
isbn = {9798400714894},
publisher = {Association for Computing Machinery},
address = {New York, NY, USA},
url = {https://doi.org/10.1145/3715928.3737482},
doi = {10.1145/3715928.3737482},
booktitle = {Proceedings of the ACM Collective Intelligence Conference},
pages = {71–82},
numpages = {12},
location = {
},
series = {CI '25}
}

@article{stahl1993evolution,
  title={Evolution of smartn players},
  author={Stahl, Dale O},
  journal={Games and Economic Behavior},
  volume={5},
  number={4},
  pages={604--617},
  year={1993},
  publisher={Elsevier}
}

@article{devaine_theory_2014,
    title = {Theory of {Mind}: {Did} {Evolution} {Fool} {Us}?},
    volume = {9},
    issn = {1932-6203},
    shorttitle = {Theory of {Mind}},
    url = {https://dx.plos.org/10.1371/journal.pone.0087619},
    doi = {10.1371/journal.pone.0087619},
    language = {en},
    number = {2},
    urldate = {2023-07-20},
    journal = {PLoS ONE},
    author = {Devaine, Marie and Hollard, Guillaume and Daunizeau, Jean},
    editor = {Zalla, Tiziana},
    month = feb,
    year = {2014},
    pages = {e87619},
}

@article{terrucha2025humans,
  title={Humans program artificial delegates to accurately solve collective-risk dilemmas but lack precision},
  author={Terrucha, In{\^e}s and Fern{\'a}ndez Domingos, Elias and Suchon, R{\'e}mi and Santos, Francisco C and Simoens, Pieter and Lenaerts, Tom},
  journal={Proceedings of the National Academy of Sciences},
  volume={122},
  number={25},
  pages={e2319942121},
  year={2025},
  publisher={National Academy of Sciences}
}

@article{demelo2018social,
  title={Social decisions and fairness change when people’s interests are represented by autonomous agents},
  author={de Melo, Celso M and Marsella, Stacy and Gratch, Jonathan},
  journal={Autonomous Agents and Multi-Agent Systems},
  volume={32},
  number={1},
  pages={163--187},
  year={2018},
  publisher={Springer}
}

@article{terrucha2024committing,
  title={Committing to the wrong artificial delegate in a collective-risk dilemma is better than directly committing mistakes},
  author={Terrucha, In{\^e}s and Fern{\'a}ndez Domingos, Elias and Simoens, Pieter and Lenaerts, Tom},
  journal={Scientific reports},
  volume={14},
  number={1},
  pages={10460},
  year={2024},
  publisher={Nature Publishing Group UK London}
}

@article{fernandez2022delegation,
  title={Delegation to artificial agents fosters prosocial behaviors in the collective risk dilemma},
  author={Fern{\'a}ndez Domingos, Elias and Terrucha, In{\^e}s and Suchon, R{\'e}mi and Gruji{\'c}, Jelena and Burguillo, Juan C and Santos, Francisco C and Lenaerts, Tom},
  journal={Scientific reports},
  volume={12},
  number={1},
  pages={8492},
  year={2022},
  publisher={Nature Publishing Group UK London}
}

@article{terrucha2024art,
  title={The art of compensation: How hybrid teams solve collective-risk dilemmas},
  author={Terrucha, In{\^e}s and Fern{\'a}ndez Domingos, Elias and C. Santos, Francisco and Simoens, Pieter and Lenaerts, Tom},
  journal={PloS one},
  volume={19},
  number={2},
  pages={e0297213},
  year={2024},
  publisher={Public Library of Science San Francisco, CA USA}
}

@article{kobis2025delegation,
  title={Delegation to artificial intelligence can increase dishonest behaviour},
  author={K{\"o}bis, Nils and Rahwan, Zoe and Rilla, Raluca and Supriyatno, Bramantyo Ibrahim and Bersch, Clara and Ajaj, Tamer and Bonnefon, Jean-Fran{\c{c}}ois and Rahwan, Iyad},
  journal={Nature},
  volume={646},
  number={8083},
  pages={126--134},
  year={2025},
  publisher={Nature Publishing Group UK London}
}

@article{griffiths2020understanding,
  title={Understanding human intelligence through human limitations},
  author={Griffiths, Thomas L},
  journal={Trends in Cognitive Sciences},
  volume={24},
  number={11},
  pages={873--883},
  year={2020},
  publisher={Elsevier}
}

@misc{bletchley_declaration_2023,
  title = {The Bletchley Declaration by Countries Attending the AI Safety Summit, 1--2 November 2023},
  author = {{AI Safety Summit}},
  year = {2023},
  month = nov,
  day = {02},
  publisher = {UK Government / GOV.UK},
  url = {https://www.gov.uk/government/publications/ai-safety-summit-2023-the-bletchley-declaration/the-bletchley-declaration-by-countries-attending-the-ai-safety-summit-1-2-november-2023},
  note = {Accessed: 02/11/2026}
}

@article{mu2024multi,
  title={Multi-agent, human--agent and beyond: a survey on cooperation in social dilemmas},
  author={Mu, Chunjiang and Guo, Hao and Chen, Yang and Shen, Chen and Hu, Die and Hu, Shuyue and Wang, Zhen},
  journal={Neurocomputing},
  volume={610},
  pages={128514},
  year={2024},
  publisher={Elsevier}
}

@techreport{bengio2026internationalAISafetyReport,
  author       = {Bengio, Yoshua and others},
  title        = {{International AI Safety Report 2026}},
  institution  = {DSIT},
  number       = {DSIT 2026/001},
  year         = {2026},
  url          = {https://internationalaisafetyreport.org},
  note         = {Accessed: 2026-02-03}
}

@article{zimmaro2024emergence,
  title={Emergence of cooperation in the one-shot Prisoner’s dilemma through Discriminatory and Samaritan AIs},
  author={Zimmaro, Filippo and Miranda, Manuel and Fern{\'a}ndez, Jos{\'e} Mar{\'\i}a Ramos and Moreno L{\'o}pez, Jes{\'u}s A and Reddel, Max and Widler, Valeria and Antonioni, Alberto and Han, The Anh},
  journal={Journal of the Royal Society Interface},
  volume={21},
  number={218},
  pages={20240212},
  year={2024},
  publisher={The Royal Society}
}

@article{hammond2025multi,
  title={Multi-agent risks from advanced ai},
  author={Hammond, Lewis and Chan, Alan and Clifton, Jesse and Hoelscher-Obermaier, Jason and Khan, Akbir and McLean, Euan and Smith, Chandler and Barfuss, Wolfram and Foerster, Jakob and Gaven{\v{c}}iak, Tom{\'a}{\v{s}} and others},
  journal={arXiv preprint arXiv:2502.14143},
  year={2025}
}

@article{wu2025minds,
  title={Minds and cooperation},
  author={Wu, Jiabin},
  journal={Rationality and Society},
  volume={37},
  number={4},
  pages={537--552},
  year={2025},
  publisher={SAGE Publications Sage UK: London, England}
}

@article{jagau2017general,
  title={A general evolutionary framework for the role of intuition and deliberation in cooperation},
  author={Jagau, Stephan and van Veelen, Matthijs},
  journal={Nature Human Behaviour},
  volume={1},
  number={8},
  pages={0152},
  year={2017},
  publisher={Nature Publishing Group UK London}
}

@article{bear2017co,
  title={Co-evolution of cooperation and cognition: the impact of imperfect deliberation and context-sensitive intuition},
  author={Bear, Adam and Kagan, Ari and Rand, David G},
  journal={Proceedings of the Royal Society B: Biological Sciences},
  volume={284},
  number={1851},
  pages={20162326},
  year={2017},
  publisher={The Royal Society}
}

@article{bear2016intuition,
  title={Intuition, deliberation, and the evolution of cooperation},
  author={Bear, Adam and Rand, David G},
  journal={Proceedings of the National Academy of Sciences},
  volume={113},
  number={4},
  pages={936--941},
  year={2016},
  publisher={National Academy of Sciences}
}

@article{couto2022introspection,
  title={Introspection dynamics: a simple model of counterfactual learning in asymmetric games},
  author={Couto, Marta C and Giaimo, Stefano and Hilbe, Christian},
  journal={New Journal of Physics},
  volume={24},
  number={6},
  pages={063010},
  year={2022},
  publisher={IOP Publishing}
}

@incollection{fernandes2024counterfactual,
  title={Counterfactual Thinking in Stochastic Dynamics of Cooperation},
  author={Fernandes, Ant{\'o}nio M and Santos, Francisco C and Paiva, Ana},
  booktitle={ECAI 2024},
  pages={3493--3500},
  year={2024},
  publisher={IOS Press}
}

@inproceedings{pereira2018counterfactual,
  title={Counterfactual thinking in cooperation dynamics},
  author={Pereira, Lu{\'\i}s Moniz and Santos, Francisco C},
  booktitle={International conference on Model-Based Reasoning},
  pages={69--82},
  year={2018},
  organization={Springer}
}

@article{lenaerts2024evolution,
  title={Evolution of a theory of mind},
  author={Lenaerts, Tom and Saponara, Marco and Pacheco, Jorge M and Santos, Francisco C},
  journal={Iscience},
  volume={27},
  number={2},
  year={2024},
  publisher={Elsevier}
}

@article{han2015synergy,
  title={Synergy between intention recognition and commitments in cooperation dilemmas},
  author={Han, The Anh and Santos, Francisco C and Lenaerts, Tom and Pereira, Luis Moniz},
  journal={Scientific reports},
  volume={5},
  number={1},
  pages={9312},
  year={2015},
  publisher={Nature Publishing Group UK London}
}

@article{han2012corpus,
  title={Corpus-based intention recognition in cooperation dilemmas},
  author={Han, The Anh and Pereira, Luis Moniz and Santos, Francisco C},
  journal={Artificial Life},
  volume={18},
  number={4},
  pages={365--383},
  year={2012},
  publisher={MIT Press 55 Hayward Street, Cambridge, MA 02142-1315, USA}
}

@article{han2011intention,
  title={Intention recognition promotes the emergence of cooperation},
  author={Han, The Anh and Moniz Pereira, Lu{\'\i}s and Santos, Francisco C},
  journal={Adaptive Behavior},
  volume={19},
  number={4},
  pages={264--279},
  year={2011},
  publisher={Sage Publications Sage UK: London, England}
}

@article{perc2025counterfeit,
  title={Counterfeit judgments in large language models},
  author={Perc, Matja{\v{z}}},
  journal={Proceedings of the National Academy of Sciences},
  volume={122},
  number={48},
  pages={e2528527122},
  year={2025},
  publisher={National Academy of Sciences}
}

@incollection{smith1982evolution,
  title={Evolution and the Theory of Games},
  author={Smith, John Maynard},
  booktitle={Did Darwin get it right? Essays on games, sex and evolution},
  pages={202--215},
  year={1982},
  publisher={Springer}
}

@article{xu2024hallucination,
  title={Hallucination is inevitable: An innate limitation of large language models},
  author={Xu, Ziwei and Jain, Sanjay and Kankanhalli, Mohan},
  journal={arXiv preprint arXiv:2401.11817},
  year={2024}
}

@inproceedings{kalai2024calibrated,
  title={Calibrated language models must hallucinate},
  author={Kalai, Adam Tauman and Vempala, Santosh S},
  booktitle={Proceedings of the 56th Annual ACM Symposium on Theory of Computing},
  pages={160--171},
  year={2024}
}

@article{price2004intuitive,
  title={Intuitive evaluation of likelihood judgment producers: Evidence for a confidence heuristic},
  author={Price, Paul C and Stone, Eric R},
  journal={Journal of Behavioral Decision Making},
  volume={17},
  number={1},
  pages={39--57},
  year={2004},
  publisher={Wiley Online Library}
}

@article{dechene2010truth,
  title={The truth about the truth: A meta-analytic review of the truth effect},
  author={Dech{\^e}ne, Alice and Stahl, Christoph and Hansen, Jochim and W{\"a}nke, Michaela},
  journal={Personality and Social Psychology Review},
  volume={14},
  number={2},
  pages={238--257},
  year={2010},
  publisher={Sage Publications Sage CA: Los Angeles, CA}
}

@article{begg1992dissociation,
  title={Dissociation of processes in belief: Source recollection, statement familiarity, and the illusion of truth.},
  author={Begg, Ian Maynard and Anas, Ann and Farinacci, Suzanne},
  journal={Journal of Experimental Psychology: General},
  volume={121},
  number={4},
  pages={446},
  year={1992},
  publisher={American Psychological Association}
}

@article{ouyang2022training,
  title={Training language models to follow instructions with human feedback},
  author={Ouyang, Long and Wu, Jeffrey and Jiang, Xu and Almeida, Diogo and Wainwright, Carroll and Mishkin, Pamela and Zhang, Chong and Agarwal, Sandhini and Slama, Katarina and Ray, Alex and others},
  journal={Advances in neural information processing systems},
  volume={35},
  pages={27730--27744},
  year={2022}
}

@article{schick2023toolformer,
  title={Toolformer: Language models can teach themselves to use tools},
  author={Schick, Timo and Dwivedi-Yu, Jane and Dess{\`\i}, Roberto and Raileanu, Roberta and Lomeli, Maria and Hambro, Eric and Zettlemoyer, Luke and Cancedda, Nicola and Scialom, Thomas},
  journal={Advances in Neural Information Processing Systems},
  volume={36},
  pages={68539--68551},
  year={2023}
}

@article{lewis2020retrieval,
  title={Retrieval-augmented generation for knowledge-intensive nlp tasks},
  author={Lewis, Patrick and Perez, Ethan and Piktus, Aleksandra and Petroni, Fabio and Karpukhin, Vladimir and Goyal, Naman and K{\"u}ttler, Heinrich and Lewis, Mike and Yih, Wen-tau and Rockt{\"a}schel, Tim and others},
  journal={Advances in neural information processing systems},
  volume={33},
  pages={9459--9474},
  year={2020}
}

@book{sandholm2010population,
  title={Population games and evolutionary dynamics},
  author={Sandholm, William H},
  year={2010},
  publisher={MIT press}
}

@book{hofbauer1998evolutionary,
  title={Evolutionary games and population dynamics},
  author={Hofbauer, Josef and Sigmund, Karl},
  year={1998},
  publisher={Cambridge university press}
}

@book{weibull1997evolutionary,
  title={Evolutionary game theory},
  author={Weibull, J{\"o}rgen W},
  year={1997},
  publisher={MIT press}
}

@article{loru2025simulation,
  title={The simulation of judgment in LLMs},
  author={Loru, Edoardo and Nudo, Jacopo and Di Marco, Niccol{\`o} and Santirocchi, Alessandro and Atzeni, Roberto and Cinelli, Matteo and Cestari, Vincenzo and Rossi-Arnaud, Clelia and Quattrociocchi, Walter},
  journal={Proceedings of the National Academy of Sciences},
  volume={122},
  number={42},
  pages={e2518443122},
  year={2025},
  publisher={National Academy of Sciences}
}

@article{quattrociocchi2025epistemological,
  title={Epistemological Fault Lines Between Human and Artificial Intelligence},
  author={Quattrociocchi, Walter and Capraro, Valerio and Perc, Matja{\v{z}}},
  journal={arXiv preprint arXiv:2512.19466},
  year={2025}
}

@article{feuerriegel2025using,
  title={Using natural language processing to analyse text data in behavioural science},
  author={Feuerriegel, Stefan and Maarouf, Abdurahman and B{\"a}r, Dominik and Geissler, Dominique and Schweisthal, Jonas and Pr{\"o}llochs, Nicolas and Robertson, Claire E and Rathje, Steve and Hartmann, Jochen and Mohammad, Saif M and others},
  journal={Nature Reviews Psychology},
  volume={4},
  number={2},
  pages={96--111},
  year={2025},
  publisher={Nature Publishing Group US New York}
}

@article{capraro2025publicly,
  title={A publicly available benchmark for assessing large language models’ ability to predict how humans balance self-interest and the interest of others},
  author={Capraro, Valerio and Di Paolo, Roberto and Pizziol, Veronica},
  journal={Scientific Reports},
  volume={15},
  number={1},
  pages={21428},
  year={2025},
  publisher={Nature Publishing Group UK London}
}

@article{bjorndahl2023sequential,
  title={Sequential language-based decisions},
  author={Bjorndahl, Adam and Halpern, Joseph Y},
  journal={arXiv preprint arXiv:2307.07563},
  year={2023}
}

@article{bjorndahl2021language,
  title={Language-based decisions},
  author={Bjorndahl, Adam and Halpern, Joseph Y},
  journal={arXiv preprint arXiv:2106.11494},
  year={2021}
}

@article{bjorndahl2013language,
  title={Language-based games},
  author={Bjorndahl, Adam and Halpern, Joseph Y and Pass, Rafael},
  journal={arXiv preprint arXiv:1310.6408},
  year={2013}
}

@article{battiston2025higher,
  title={Higher-order interactions shape collective human behaviour},
  author={Battiston, Federico and Capraro, Valerio and Karimi, Fariba and Lehmann, Sune and Migliano, Andrea Bamberg and Sadekar, Onkar and S{\'a}nchez, Angel and Perc, Matja{\v{z}}},
  journal={Nature Human Behaviour},
  pages={1--17},
  year={2025},
  publisher={Nature Publishing Group UK London}
}

@article{mei2024turing,
  title={A Turing test of whether AI chatbots are behaviorally similar to humans},
  author={Mei, Qiaozhu and Xie, Yutong and Yuan, Walter and Jackson, Matthew O},
  journal={Proceedings of the National Academy of Sciences},
  volume={121},
  number={9},
  pages={e2313925121},
  year={2024},
  publisher={National Acad Sciences}
}

@techreport{horton2023large,
  title={Large language models as simulated economic agents: What can we learn from homo silicus?},
  author={Horton, John J},
  year={2023},
  institution={National Bureau of Economic Research}
}

@article{dillion2023can,
  title={Can AI language models replace human participants?},
  author={Dillion, Danica and Tandon, Niket and Gu, Yuling and Gray, Kurt},
  journal={Trends in Cognitive Sciences},
  volume={27},
  number={7},
  pages={597--600},
  year={2023},
  publisher={Elsevier}
}

@article{chen2023emergence,
  title={The emergence of economic rationality of GPT},
  author={Chen, Yiting and Liu, Tracy Xiao and Shan, You and Zhong, Songfa},
  journal={Proceedings of the National Academy of Sciences},
  volume={120},
  number={51},
  pages={e2316205120},
  year={2023},
  publisher={National Acad Sciences}
}

@article{alger2024doing,
  title={Doing the right thing (or not) in a lemons-like situation: on the role of social preferences and Kantian moral concerns},
  author={Alger, Ingela and Rivero-Wildemauwe, Jos{\'e} Ignacio},
  journal={arXiv preprint arXiv:2405.13186},
  year={2024}
}

@article{jimenez2023conditioning,
  title={Conditioning competitive behaviour in experimental Bertrand markets through contextual frames},
  author={Jim{\'e}nez-Jim{\'e}nez, Francisca and Rodero-Cosano, Javier},
  journal={Journal of Behavioral and Experimental Economics},
  volume={103},
  pages={101987},
  year={2023},
  publisher={Elsevier}
}

@article{scigala2022corrupting,
  title={Corrupting the prosocial people: does cooperation framing increase bribery engagement among prosocial individuals? Stage 1 Registered Report},
  author={{\'S}ciga{\l}a, Karolina Aleksandra and Zettler, Ingo and Pfattheicher, Stefan and Capraro, Valerio},
  year={2022},
  publisher={PsyArXiv}
}

@article{kuang2024language,
  title={Language matters: how normative expressions shape norm perception and affect norm compliance},
  author={Kuang, Jinyi and Bicchieri, Cristina},
  journal={Philosophical Transactions of the Royal Society B},
  volume={379},
  number={1897},
  pages={20230037},
  year={2024},
  publisher={The Royal Society}
}

@article{chang2019rhetoric,
  title={Rhetoric matters: A social norms explanation for the anomaly of framing},
  author={Chang, Daphne and Chen, Roy and Krupka, Erin},
  journal={Games and Economic Behavior},
  volume={116},
  pages={158--178},
  year={2019},
  publisher={Elsevier}
}

@article{capraro2019power,
  title={The power of moral words: Loaded language generates framing effects in the extreme dictator game},
  author={Capraro, Valerio and Vanzo, Andrea},
  journal={Judgment and Decision Making},
  volume={14},
  number={3},
  pages={309--317},
  year={2019},
  publisher={Cambridge University Press}
}

@article{huang2019choosing,
  title={Choosing an equitable or efficient option: A distribution dilemma},
  author={Huang, Long and Lei, Wansheng and Xu, Fuming and Yu, Liang and Shi, Fujun},
  journal={Social Behavior and Personality: an international journal},
  volume={47},
  number={10},
  pages={1--10},
  year={2019},
  publisher={Scientific Journal Publishers}
}

@article{huang2020maxims,
  title={Maxims nudge equitable or efficient choices in a Trade-Off Game},
  author={Huang, Long and Lei, Wansheng and Xu, Fuming and Liu, Hairong and Yu, Liang and Shi, Fujun and Wang, Lei},
  journal={PloS one},
  volume={15},
  number={6},
  pages={e0235443},
  year={2020},
  publisher={Public Library of Science San Francisco, CA USA}
}

@article{capraro2018right,
  title={Do the right thing: Experimental evidence that preferences for moral behavior, rather than equity or efficiency per se, drive human prosociality},
  author={Capraro, Valerio and Rand, David G},
  journal={Judgment and Decision Making},
  volume={13},
  number={1},
  pages={99--111},
  year={2018},
  publisher={Cambridge University Press}
}

@article{eriksson2017costly,
  title={Costly punishment in the ultimatum game evokes moral concern, in particular when framed as payoff reduction},
  author={Eriksson, Kimmo and Strimling, Pontus and Andersson, Per A and Lindholm, Torun},
  journal={Journal of Experimental Social Psychology},
  volume={69},
  pages={59--64},
  year={2017},
  publisher={Elsevier}
}

@article{liberman2004name,
  title={The name of the game: Predictive power of reputations versus situational labels in determining prisoner’s dilemma game moves},
  author={Liberman, Varda and Samuels, Steven M and Ross, Lee},
  journal={Personality and social psychology bulletin},
  volume={30},
  number={9},
  pages={1175--1185},
  year={2004},
  publisher={Sage Publications Sage CA: Thousand Oaks, CA}
}

@article{mieth2021moral,
  title={Moral labels increase cooperation and costly punishment in a Prisoner’s Dilemma game with punishment option},
  author={Mieth, Laura and Buchner, Axel and Bell, Raoul},
  journal={Scientific Reports},
  volume={11},
  number={1},
  pages={10221},
  year={2021},
  publisher={Nature Publishing Group UK London}
}

@article{engel2014does,
  title={What does “clean” really mean? The implicit framing of decontextualized experiments},
  author={Engel, Christoph and Rand, David G},
  journal={Economics Letters},
  volume={122},
  number={3},
  pages={386--389},
  year={2014},
  publisher={Elsevier}
}

@article{perrault2024artificial,
  title={Artificial Intelligence Index Report 2024},
  author={Perrault, Ray and Clark, Jack},
  year={2024}
}

@article{zhang2021ethics,
  title={Ethics and governance of artificial intelligence: Evidence from a survey of machine learning researchers},
  author={Zhang, Baobao and Anderljung, Markus and Kahn, Lauren and Dreksler, Noemi and Horowitz, Michael C and Dafoe, Allan},
  journal={Journal of Artificial Intelligence Research},
  volume={71},
  pages={591--666},
  year={2021}
}

@article{capraro2024language,
  title={Language-based game theory in the age of artificial intelligence},
  author={Capraro, Valerio and Di Paolo, Roberto and Perc, Matja{\v{z}} and Pizziol, Veronica},
  journal={Journal of the Royal Society Interface},
  volume={21},
  number={212},
  pages={20230720},
  year={2024},
  publisher={The Royal Society}
}

@article{roberts2023governing,
  title={Governing artificial intelligence in China and the European Union: Comparing aims and promoting ethical outcomes},
  author={Roberts, Huw and Cowls, Josh and Hine, Emmie and Morley, Jessica and Wang, Vincent and Taddeo, Mariarosaria and Floridi, Luciano},
  journal={The Information Society},
  volume={39},
  number={2},
  pages={79--97},
  year={2023},
  publisher={Taylor \& Francis}
}

@article{pamlin2015global,
  title={Global challenges: 12 risks that threaten human civilization},
  author={Pamlin, Dennis and Armstrong, Stuart},
  journal={Global Challenges Foundation, Stockholm},
  year={2015}
}

@book{ord2020precipice,
  title={The precipice: Existential risk and the future of humanity},
  author={Ord, Toby},
  year={2020},
  publisher={Hachette Books}
}

@article{baliga2004arms,
  title={Arms races and negotiations},
  author={Baliga, Sandeep and Sj{\"o}str{\"o}m, Tomas},
  journal={The Review of Economic Studies},
  volume={71},
  number={2},
  pages={351--369},
  year={2004},
  publisher={Wiley-Blackwell}
}

@article{santos2011risk,
  title={Risk of collective failure provides an escape from the tragedy of the commons},
  author={Santos, Francisco C and Pacheco, Jorge M},
  journal={Proceedings of the National Academy of Sciences},
  volume={108},
  number={26},
  pages={10421--10425},
  year={2011},
  publisher={National Acad Sciences}
}

@article{milinski2008collective,
  title={The collective-risk social dilemma and the prevention of simulated dangerous climate change},
  author={Milinski, Manfred and Sommerfeld, Ralf D and Krambeck, Hans-J{\"u}rgen and Reed, Floyd A and Marotzke, Jochem},
  journal={Proceedings of the National Academy of Sciences},
  volume={105},
  number={7},
  pages={2291--2294},
  year={2008},
  publisher={National Acad Sciences}
}

@inproceedings{zhan2023deceptive,
  title={Deceptive AI ecosystems: The case of ChatGPT},
  author={Zhan, Xiao and Xu, Yifan and Sarkadi, Stefan},
  booktitle={Proceedings of the 5th International Conference on Conversational User Interfaces},
  pages={1--6},
  year={2023}
}

@article{alalawi2024trust,
title = {Trust AI regulation? Discerning users are vital to build trust and effective AI regulation},
journal = {Applied Mathematics and Computation},
volume = {508},
pages = {129627},
year = {2026},
issn = {0096-3003},
doi = {https://doi.org/10.1016/j.amc.2025.129627},
url = {https://www.sciencedirect.com/science/article/pii/S0096300325003534},
author = {Zainab Alalawi and Paolo Bova and Theodor Cimpeanu and Alessandro {Di Stefano} and Manh {Hong Duong} and Elias Fernández Domingos and The Anh Han and Marcus Krellner and Ndidi Bianca Ogbo and Simon T. Powers and Filippo Zimmaro}
}

@article{balabanova2025media,
  title={Media and responsible AI governance: a game-theoretic and LLM analysis},
  author={Balabanova, Nataliya and Bashir, Adeela and Bova, Paolo and Buscemi, Alessio and Cimpeanu, Theodor and da Fonseca, Henrique Correia and Di Stefano, Alessandro and Duong, Manh Hong and Domingos, Elias Fernandez and Fernandes, Antonio and others},
  journal={arXiv preprint arXiv:2503.09858},
  year={2025}
}

@article{chang2024survey,
  title={A survey on evaluation of large language models},
  author={Chang, Yupeng and Wang, Xu and Wang, Jindong and Wu, Yuan and Yang, Linyi and Zhu, Kaijie and Chen, Hao and Yi, Xiaoyuan and Wang, Cunxiang and Wang, Yidong and others},
  journal={ACM transactions on intelligent systems and technology},
  volume={15},
  number={3},
  pages={1--45},
  year={2024},
  publisher={ACM New York, NY}
}

@inproceedings{han2019modelling,
  title={Modelling and influencing the AI bidding war: a research agenda},
  author={Han, The Anh and Pereira, Lu{\'\i}s Moniz and Lenaerts, Tom},
  booktitle={Proceedings of the 2019 AAAI/ACM Conference on AI, Ethics, and Society},
  pages={5--11},
  year={2019}
}

@article{bova2024both,
  title={Both eyes open: Vigilant incentives help auditors improve ai safety},
  author={Bova, Paolo and Stefano, Alessandro Di and Han, The Anh},
  journal={Journal of Physics: Complexity},
  volume={5},
  number={2},
  pages={025009},
  year={2024},
  publisher={IOP Publishing}
}

@inproceedings{o2020windfall,
  title={The windfall clause: Distributing the benefits of AI for the common good},
  author={O'Keefe, Cullen and Cihon, Peter and Garfinkel, Ben and Flynn, Carrick and Leung, Jade and Dafoe, Allan},
  booktitle={Proceedings of the AAAI/ACM Conference on AI, Ethics, and Society},
  pages={327--331},
  year={2020}
}

@misc{dignum2025roadmap,
  title={Roadmap for AI policy research: AI policy research summit, Stockholm, November 2024},
  author={Dignum, Virginia and R{\'e}gis, Catherine and Bach, Kerstin and PLF de Carvalho, Andr{\'e} and Castellano, Ginevra and Dignum, Frank and Farries, Elizabeth and Giannotti, Fosca and Helberger, Natali and Hellegren, Isadora and others},
  year={2025}
}

@article{emery2023uncertainty,
  title={Uncertainty, information, and risk in international technology races},
  author={Emery-Xu, Nicholas and Park, Andrew and Trager, Robert},
  journal={Journal of Conflict Resolution},
  pages={00220027231214996},
  year={2023},
  publisher={SAGE Publications Sage CA: Los Angeles, CA}
}

@article{dafoe2021cooperative,
  title={Cooperative AI: machines must learn to find common ground},
  author={Dafoe, Allan and Bachrach, Yoram and Hadfield, Gillian and Horvitz, Eric and Larson, Kate and Graepel, Thore},
  year={2021},
  publisher={Nature Publishing Group}
}

@article{santos2024prosocial,
  title={Prosocial dynamics in multiagent systems},
  author={Santos, Fernando P},
  journal={AI Magazine},
  volume={45},
  number={1},
  pages={131--138},
  year={2024},
  publisher={Wiley Online Library}
}

@article{capraro2024impact,
  title={The impact of generative artificial intelligence on socioeconomic inequalities and policy making},
  author={Capraro, Valerio and Lentsch, Austin and Acemoglu, Daron and Akgun, Selin and Akhmedova, Aisel and Bilancini, Ennio and Bonnefon, Jean-Fran{\c{c}}ois and Bra{\~n}as-Garza, Pablo and Butera, Luigi and Douglas, Karen M and others},
  journal={PNAS nexus},
  volume={3},
  number={6},
  year={2024},
  publisher={Oxford Academic}
}

@article{mclean2023risks,
  title={The risks associated with Artificial General Intelligence: A systematic review},
  author={McLean, Scott and Read, Gemma JM and Thompson, Jason and Baber, Chris and Stanton, Neville A and Salmon, Paul M},
  journal={Journal of Experimental \& Theoretical Artificial Intelligence},
  volume={35},
  number={5},
  pages={649--663},
  year={2023},
  publisher={Taylor \& Francis}
}

@article{bengio2024managing,
  title={Managing extreme AI risks amid rapid progress},
  author={Bengio, Yoshua and Hinton, Geoffrey and Yao, Andrew and Song, Dawn and Abbeel, Pieter and Darrell, Trevor and Harari, Yuval Noah and Zhang, Ya-Qin and Xue, Lan and Shalev-Shwartz, Shai and others},
  journal={Science},
  volume={384},
  number={6698},
  pages={842--845},
  year={2024},
  publisher={American Association for the Advancement of Science}
}

@article{sharma2023small,
  title={Small bots, big impact: solving the conundrum of cooperation in optional Prisoner’s Dilemma game through simple strategies},
  author={Sharma, Gopal and Guo, Hao and Shen, Chen and Tanimoto, Jun},
  journal={Journal of The Royal Society Interface},
  volume={20},
  number={204},
  pages={20230301},
  year={2023},
  publisher={The Royal Society}
}

@article{veselovsky2023artificial,
  title={Artificial artificial artificial intelligence: Crowd workers widely use large language models for text production tasks},
  author={Veselovsky, Veniamin and Ribeiro, Manoel Horta and West, Robert},
  journal={arXiv preprint arXiv:2306.07899},
  year={2023}
}

@article{shumailov2023curse,
  title={The curse of recursion: Training on generated data makes models forget},
  author={Shumailov, Ilia and Shumaylov, Zakhar and Zhao, Yiren and Gal, Yarin and Papernot, Nicolas and Anderson, Ross},
  journal={arXiv preprint arXiv:2305.17493},
  year={2023}
}

@article{mason2012collaborative,
  title={Collaborative learning in networks},
  author={Mason, Winter and Watts, Duncan J},
  journal={Proceedings of the National Academy of Sciences},
  volume={109},
  number={3},
  pages={764--769},
  year={2012},
  publisher={National Acad Sciences}
}

@article{lazer2007network,
  title={The network structure of exploration and exploitation},
  author={Lazer, David and Friedman, Allan},
  journal={Administrative science quarterly},
  volume={52},
  number={4},
  pages={667--694},
  year={2007},
  publisher={SAGE Publications Sage CA: Los Angeles, CA}
}

@inproceedings{su2016effect,
  title={The effect of recommendations on network structure},
  author={Su, Jessica and Sharma, Aneesh and Goel, Sharad},
  booktitle={Proceedings of the 25th international conference on World Wide Web},
  pages={1157--1167},
  year={2016}
}

@article{lucas2020value,
  title={The value of teaching increases with tool complexity in cumulative cultural evolution},
  author={Lucas, Amanda J and Kings, Michael and Whittle, Devi and Davey, Emma and Happ{\'e}, Francesca and Caldwell, Christine A and Thornton, Alex},
  journal={Proceedings of the Royal Society B},
  volume={287},
  number={1939},
  pages={20201885},
  year={2020},
  publisher={The Royal Society}
}

@article{li2017survey,
  title={A survey of link recommendation for social networks: Methods, theoretical foundations, and future research directions},
  author={Li, Zhepeng and Fang, Xiao and Sheng, Olivia R Liu},
  journal={ACM Transactions on Management Information Systems (TMIS)},
  volume={9},
  number={1},
  pages={1--26},
  year={2017},
  publisher={ACM New York, NY, USA}
}

@inproceedings{baumann2024optimal,
  title={Optimal Engagement-Diversity Tradeoffs in Social Media},
  author={Baumann, Fabian and Halpern, Daniel and Procaccia, Ariel D and Rahwan, Iyad and Shapira, Itai and W{\"u}thrich, Manuel},
  booktitle={Proceedings of the ACM on Web Conference 2024},
  pages={288--299},
  year={2024}
}

@inproceedings{rombach2022high,
  title={High-resolution image synthesis with latent diffusion models},
  author={Rombach, Robin and Blattmann, Andreas and Lorenz, Dominik and Esser, Patrick and Ommer, Bj{\"o}rn},
  booktitle={Proceedings of the IEEE/CVF conference on computer vision and pattern recognition},
  pages={10684--10695},
  year={2022}
}

@article{johnson2021openai,
  title={OpenAI debuts DALL-E for generating images from text},
  author={Johnson, Khari},
  journal={VentureBeat},
  year={2021}
}

@article{mesoudi2016cultural,
  title={Cultural evolution: a review of theory, findings and controversies},
  author={Mesoudi, Alex},
  journal={Evolutionary biology},
  volume={43},
  pages={481--497},
  year={2016},
  publisher={Springer}
}

@article{capraro2024outcome,
  title={From outcome-based to language-based preferences},
  author={Capraro, Valerio and Halpern, Joseph Y and Perc, Matja{\v{z}}},
  journal={Journal of Economic Literature},
  volume={62},
  number={1},
  pages={115--154},
  year={2024},
  publisher={American Economic Association 2014 Broadway, Suite 305, Nashville, TN 37203-2425}
}

@article{brinkmann2023machine,
  title={Machine culture},
  author={Brinkmann, Levin and Baumann, Fabian and Bonnefon, Jean-Fran{\c{c}}ois and Derex, Maxime and M{\"u}ller, Thomas F and Nussberger, Anne-Marie and Czaplicka, Agnieszka and Acerbi, Alberto and Griffiths, Thomas L and Henrich, Joseph and others},
  journal={Nature Human Behaviour},
  volume={7},
  number={11},
  pages={1855--1868},
  year={2023},
  publisher={Nature Publishing Group UK London}
}

@article{rahwan2019machine,
  title={Machine behaviour},
  author={Rahwan, Iyad and Cebrian, Manuel and Obradovich, Nick and Bongard, Josh and Bonnefon, Jean-Fran{\c{c}}ois and Breazeal, Cynthia and Crandall, Jacob W and Christakis, Nicholas A and Couzin, Iain D and Jackson, Matthew O and others},
  journal={Nature},
  volume={568},
  number={7753},
  pages={477--486},
  year={2019},
  publisher={Nature Publishing Group UK London}
}

@article{andras2018trusting,
  title={Trusting intelligent machines: Deepening trust within socio-technical systems},
  author={Andras, Peter and Esterle, Lukas and Guckert, Michael and Han, The Anh and Lewis, Peter R and Milanovic, Kristina and Payne, Terry and Perret, Cedric and Pitt, Jeremy and Powers, Simon T and others},
  journal={IEEE Technology and Society Magazine},
  volume={37},
  number={4},
  pages={76--83},
  year={2018},
  publisher={IEEE}
}

@article{piao2023human,
  title={Human--AI adaptive dynamics drives the emergence of information cocoons},
  author={Piao, Jinghua and Liu, Jiazhen and Zhang, Fang and Su, Jun and Li, Yong},
  journal={Nature Machine Intelligence},
  volume={5},
  number={11},
  pages={1214--1224},
  year={2023},
  publisher={Nature Publishing Group UK London}
}

@article{han2021or,
  title={When to (or not to) trust intelligent machines: Insights from an evolutionary game theory analysis of trust in repeated games},
  author={Han, The Anh and Perret, Cedric and Powers, Simon T},
  journal={Cognitive Systems Research},
  volume={68},
  pages={111--124},
  year={2021},
  publisher={Elsevier}
}

@book{clark2023meaningful,
  title={Meaningful games: Exploring language with game theory},
  author={Clark, Robin},
  year={2023},
  publisher={MIT Press}
}

@article{rathje2024gpt,
  title={GPT is an effective tool for multilingual psychological text analysis},
  author={Rathje, Steve and Mirea, Dan-Mircea and Sucholutsky, Ilia and Marjieh, Raja and Robertson, Claire E and Van Bavel, Jay J},
  journal={Proceedings of the National Academy of Sciences},
  volume={121},
  number={34},
  pages={e2308950121},
  year={2024},
  publisher={National Academy of Sciences}
}

@article{curry2019good,
  title={Is it good to cooperate? Testing the theory of morality-as-cooperation in 60 societies},
  author={Curry, Oliver Scott and Mullins, Daniel Austin and Whitehouse, Harvey},
  journal={Current anthropology},
  volume={60},
  number={1},
  pages={47--69},
  year={2019},
  publisher={The University of Chicago Press Chicago, IL}
}

@incollection{gaffal2024negotiation,
  title={Negotiation, Game Theory and Language Games},
  author={Gaffal, Margit and Padilla G{\'a}lvez, Jes{\'u}s},
  booktitle={Dynamics of Rational Negotiation: Game Theory, Language Games and Forms of Life},
  pages={11--40},
  year={2024},
  publisher={Springer}
}

@article{danovski2022evolutionary,
  title={On the evolutionary language game in structured and adaptive populations},
  author={Danovski, Kaloyan and Brede, Markus},
  journal={Plos one},
  volume={17},
  number={8},
  pages={e0273608},
  year={2022},
  publisher={Public Library of Science San Francisco, CA USA}
}

@article{baronchelli2006sharp,
  title={Sharp transition towards shared vocabularies in multi-agent systems},
  author={Baronchelli, Andrea and Felici, Maddalena and Loreto, Vittorio and Caglioti, Emanuele and Steels, Luc},
  journal={Journal of Statistical Mechanics: Theory and Experiment},
  volume={2006},
  number={06},
  pages={P06014},
  year={2006},
  publisher={IOP Publishing}
}

@article{puglisi2008cultural,
  title={Cultural route to the emergence of linguistic categories},
  author={Puglisi, Andrea and Baronchelli, Andrea and Loreto, Vittorio},
  journal={Proceedings of the National Academy of Sciences},
  volume={105},
  number={23},
  pages={7936--7940},
  year={2008},
  publisher={National Acad Sciences}
}

@article{kumar2020evolution,
  title={The evolution of trust and trustworthiness},
  author={Kumar, Aanjaneya and Capraro, Valerio and Perc, Matja{\v{z}}},
  journal={Journal of the Royal Society Interface},
  volume={17},
  number={169},
  pages={20200491},
  year={2020},
  publisher={The Royal Society}
}

@article{capraro2019evolution,
  title={The evolution of lying in well-mixed populations},
  author={Capraro, Valerio and Perc, Matja{\v{z}} and Vilone, Daniele},
  journal={Journal of the Royal Society Interface},
  volume={16},
  number={156},
  pages={20190211},
  year={2019},
  publisher={The Royal Society}
}

@article{perc2017statistical,
  title={Statistical physics of human cooperation},
  author={Perc, Matja{\v{z}} and Jordan, Jillian J and Rand, David G and Wang, Zhen and Boccaletti, Stefano and Szolnoki, Attila},
  journal={Physics Reports},
  volume={687},
  pages={1--51},
  year={2017},
  publisher={Elsevier}
}

@article{nowak2006five,
  title={Five rules for the evolution of cooperation},
  author={Nowak, Martin A},
  journal={science},
  volume={314},
  number={5805},
  pages={1560--1563},
  year={2006},
  publisher={American Association for the Advancement of Science}
}

@book{frank88,
	Author = {Frank, Robert H.},
	Publisher = {Norton and Company},
	Title = {Passions {W}ithin {R}eason: {T}he {S}trategic {R}ole of the {E}motions},
	Year = {1988}}

@article{armstrong2016racing,
  title        = {{   Racing to the Precipice: A Model of Artificial Intelligence Development  }},
  shorttitle   = {Racing to the Precipice},
  author       = {Armstrong, Stuart and Bostrom, Nick and others},
  year         = 2016,
  month        = may,
  journal      = {Ai \& Society},
  volume       = 31,
  number       = 2,
  pages        = {201--206},
  doi          = {10.1007/s00146-015-0590-y},
  issn         = {1435-5655},
  langid       = {english}
}

@article{cimpeanu2022artificial,
  title        = {{Artificial Intelligence Development Races in Heterogeneous Settings}},
  author       = {Cimpeanu, Theodor and Santos, Francisco and others},
  year         = 2022,
  journal      = {Scientific Reports},
  publisher    = {{Springer Science and Business Media LLC}},
  volume       = 12,
  number       = 1,
  pages        = 1723,
  doi          = {10.1038/s41598-022-05729-3}
}

@misc{hadfield2023regulatory,
  title        = {{Regulatory Markets: The Future of AI Governance}},
  shorttitle   = {Regulatory {{Markets}}},
  author       = {Hadfield, Gillian K. and Clark, Jack},
  year         = 2023,
  month        = apr,
  publisher    = {arXiv},
  number       = {arXiv:2304.04914},
  doi          = {10.48550/arXiv.2304.04914},
  urldate      = {2023-04-18},
  eprint       = {2304.04914},
  primaryclass = {cs, econ, q-fin},
  archiveprefix = {arxiv}
}

@article{han2020regulate,
  title        = {{ To Regulate or Not: A Social Dynamics Analysis of an Idealised AI Race }},
  shorttitle   = {To {{Regulate}} or {{Not}}},
  author       = {Han, The Anh and Pereira, Luis Moniz and others},
  year         = 2020,
  month        = nov,
  journal      = {Journal of Artificial Intelligence Research},
  volume       = 69,
  pages        = {881--921},
  doi          = {10.1613/jair.1.12225},
  issn         = {1076-9757},
  copyright    = {Copyright (c) 2020 Journal of Artificial Intelligence Research},
  langid       = {english}
}

@article{han2021mediating,
  title        = {{   Mediating Artificial Intelligence Developments through Negative and Positive Incentives   }},
  author       = {Han, The Anh and Pereira, Lu{\'i}s Moniz and others},
  year         = 2021,
  month        = jan,
  journal      = {Plos One},
  publisher    = {Public Library of Science},
  volume       = 16,
  number       = 1,
  doi          = {10.1371/journal.pone.0244592},
  issn         = {1932-6203},
  langid       = {english}
}

@article{han2022voluntary,
  title        = {{   Voluntary Safety Commitments Provide an Escape from Over-Regulation in AI Development   }},
  author       = {Han, The Anh and Lenaerts, Tom and others},
  year         = 2022,
  journal      = {Technology in Society},
  publisher    = {Elsevier},
  volume       = 68,
  pages        = 101843
}

@misc{maslej2023ai,
  title        = {{The AI Index 2023 Annual Report}},
  author       = {Maslej, Nestor and Fattorini, Loredana and others},
  year         = 2023,
  urldate      = {2023-03-03}
}

@article{powers2023stuff,
  title        = {{The Stuff We Swim in: Regulation Alone Will Not Lead to Justifiable Trust in AI}},
  author       = {Powers, Simon T and Linnyk, Olena and others},
  year         = 2023,
  journal      = {IEEE Technology and Society Magazine},
  publisher    = {Ieee},
  volume       = 42,
  number       = 4,
  pages        = {95--106}
}

@article{mckee2023humans,
  title={Humans perceive warmth and competence in artificial intelligence},
  author={McKee, Kevin R and Bai, Xuechunzi and Fiske, Susan T},
  journal={Iscience},
  volume={26},
  number={8},
  year={2023},
  publisher={Elsevier}
}

@article{bergman2024stela,
  title={STELA: a community-centred approach to norm elicitation for AI alignment},
  author={Bergman, Stevie and Marchal, Nahema and Mellor, John and Mohamed, Shakir and Gabriel, Iason and Isaac, William},
  journal={Scientific Reports},
  volume={14},
  number={1},
  pages={6616},
  year={2024},
  publisher={Nature Publishing Group UK London}
}

@article{duenez2023social,
  title={A social path to human-like artificial intelligence},
  author={Du{\'e}{\~n}ez-Guzm{\'a}n, Edgar A and Sadedin, Suzanne and Wang, Jane X and McKee, Kevin R and Leibo, Joel Z},
  journal={Nature Machine Intelligence},
  volume={5},
  number={11},
  pages={1181--1188},
  year={2023},
  publisher={Nature Publishing Group UK London}
}

@article{leibo2019autocurricula,
  title={Autocurricula and the emergence of innovation from social interaction: A manifesto for multi-agent intelligence research},
  author={Leibo, Joel Z and Hughes, Edward and Lanctot, Marc and Graepel, Thore},
  journal={arXiv preprint arXiv:1903.00742},
  year={2019}
}

@inproceedings{smith2025evaluating,
  title={Evaluating generalization capabilities of LLM-based agents in mixed-motive scenarios using concordia},
  author={Smith, Chandler and Abdulhai, Marwa and Diaz, Manfred and Tesic, Marko and Trivedi, Rakshit S and Vezhnevets, Alexander Sasha and Hammond, Lewis and Clifton, Jesse and Chang, Minsuk and Du{\'e}{\~n}ez-Guzm{\'a}n, Edgar A and others},
  booktitle={NeurIPS 2024 Competition Track},
  year={2025}
}

@article{vezhnevets2023generative,
  title={Generative agent-based modeling with actions grounded in physical, social, or digital space using Concordia},
  author={Vezhnevets, Alexander Sasha and Agapiou, John P and Aharon, Avia and Ziv, Ron and Matyas, Jayd and Du{\'e}{\~n}ez-Guzm{\'a}n, Edgar A and Cunningham, William A and Osindero, Simon and Karmon, Danny and Leibo, Joel Z},
  journal={arXiv preprint arXiv:2312.03664},
  year={2023}
}

@article{leibo2024theory,
  title={A theory of appropriateness with applications to generative artificial intelligence},
  author={Leibo, Joel Z and Vezhnevets, Alexander Sasha and Diaz, Manfred and Agapiou, John P and Cunningham, William A and Sunehag, Peter and Haas, Julia and Koster, Raphael and Du{\'e}{\~n}ez-Guzm{\'a}n, Edgar A and Isaac, William S and Piliouras, Georgios and Bileschi, Stan M and Rahwan, Iyad and Osindero, Simon},
  journal={arXiv preprint arXiv:2412.19010},
  year={2024}
}

@book{diresta2024invisible,
  title={Invisible rulers: The people who turn lies into reality},
  author={DiResta, Renee},
  year={2024},
  publisher={PublicAffairs}
}

@article{cohen2024regulating,
  title={Regulating advanced artificial agents},
  author={Cohen, Michael K and Kolt, Noam and Bengio, Yoshua and Hadfield, Gillian K and Russell, Stuart},
  journal={Science},
  volume={384},
  number={6691},
  pages={36--38},
  year={2024},
  publisher={American Association for the Advancement of Science}
}

@article{tomasev2025virtual,
  title={Virtual agent economies},
  author={Tomasev, Nenad and Franklin, Matija and Leibo, Joel Z and Jacobs, Julian and Cunningham, William A and Gabriel, Iason and Osindero, Simon},
  journal={arXiv preprint arXiv:2509.10147},
  year={2025}
}

@article{hadfield2025economy,
  title={An economy of ai agents},
  author={Hadfield, Gillian K and Koh, Andrew},
  journal={arXiv preprint arXiv:2509.01063},
  year={2025},
  publisher={arXiv}
}

@article{leibo2025societal,
  title={Societal and technological progress as sewing an ever-growing, ever-changing, patchy, and polychrome quilt},
  author={Leibo, Joel Z and Vezhnevets, Alexander Sasha and Cunningham, William A and Krier, S{\'e}bastien and Diaz, Manfred and Osindero, Simon},
  journal={arXiv preprint arXiv:2505.05197},
  year={2025}
}

@article{gabriel2020artificial,
  title={Artificial intelligence, values, and alignment},
  author={Gabriel, Iason},
  journal={Minds and machines},
  volume={30},
  number={3},
  pages={411--437},
  year={2020},
  publisher={Springer}
}

@article{dafoe2020open,
  title={Open problems in cooperative AI},
  author={Dafoe, Allan and Hughes, Edward and Bachrach, Yoram and Collins, Tantum and McKee, Kevin R and Leibo, Joel Z and Larson, Kate and Graepel, Thore},
  journal={arXiv preprint arXiv:2012.08630},
  year={2020}
}

@article{gabriel2024ethics,
  title={The ethics of advanced ai assistants},
  author={Gabriel, Iason and Manzini, Arianna and Keeling, Geoff and Hendricks, Lisa Anne and Rieser, Verena and Iqbal, Hasan and Toma{\v{s}}ev, Nenad and Ktena, Ira and Kenton, Zachary and Rodriguez, Mikel and others},
  journal={arXiv preprint arXiv:2404.16244},
  year={2024}
}

@article{huang2023generative,
  title={Generative AI and the digital commons},
  author={Huang, Saffron and Siddarth, Divya},
  journal={arXiv preprint arXiv:2303.11074},
  year={2023}
}

@incollection{lugrin2021introduction,
  title={Introduction to socially interactive agents},
  author={Lugrin, Birgit},
  booktitle={The handbook on socially interactive agents: 20 years of research on embodied conversational agents, intelligent virtual agents, and social robotics volume 1: methods, behavior, cognition},
  pages={1--20},
  year={2021}
}

@inproceedings{santos2019evolution,
  title={Evolution of collective fairness in hybrid populations of humans and agents},
  author={Santos, Fernando P and Pacheco, Jorge M and Paiva, Ana and Santos, Francisco C},
  booktitle={Proceedings of the AAAI conference on artificial intelligence},
  volume={33},
  number={01},
  pages={6146--6153},
  year={2019}
}

@article{guo2023facilitating,
  title={Facilitating cooperation in human-agent hybrid populations through autonomous agents},
  author={Guo, Hao and Shen, Chen and Hu, Shuyue and Xing, Junliang and Tao, Pin and Shi, Yuanchun and Wang, Zhen},
  journal={iScience},
  volume={26},
  number={11},
  year={2023},
  publisher={Elsevier}
}

@article{shirado2017locally,
  title={Locally noisy autonomous agents improve global human coordination in network experiments},
  author={Shirado, Hirokazu and Christakis, Nicholas A},
  journal={Nature},
  volume={545},
  number={7654},
  pages={370--374},
  year={2017},
  publisher={Nature Publishing Group UK London}
}

@inproceedings{da2025can,
  title={Can Media Act as a Soft Regulator of Safe AI Development? A Game Theoretical Analysis},
  author={Da Fonseca, Henrique Correia and Fernandes, Ant{\'o}nio and Song, Zhao and Cimpeanu, Theodor and Balabanova, Nataliya and Bashir, Adeela and Bova, Paolo and Buscemi, Alessio and Di Stefano, Alessandro and Duong, Manh Hong and others},
  booktitle={Artificial Life Conference Proceedings 37},
  volume={2025},
  number={1},
  pages={90},
  year={2025},
  organization={MIT Press}
}

@article{bonnefon2024moral,
  title={The moral psychology of artificial intelligence},
  author={Bonnefon, Jean-Fran{\c{c}}ois and Rahwan, Iyad and Shariff, Azim},
  journal={Annual review of psychology},
  volume={75},
  number={1},
  pages={653--675},
  year={2024},
  publisher={Annual Reviews}
}

@article{lu2024llms,
  title={LLMs and generative agent-based models for complex systems research},
  author={Lu, Yikang and Aleta, Alberto and Du, Chunpeng and Shi, Lei and Moreno, Yamir},
  journal={Physics of Life Reviews},
  volume={51},
  pages={283--293},
  year={2024},
  publisher={Elsevier}
}

@article{pedreschi2025human,
  title={Human-AI coevolution},
  author={Pedreschi, Dino and Pappalardo, Luca and Ferragina, Emanuele and Baeza-Yates, Ricardo and Barab{\'a}si, Albert-L{\'a}szl{\'o} and Dignum, Frank and Dignum, Virginia and Eliassi-Rad, Tina and Giannotti, Fosca and Kert{\'e}sz, J{\'a}nos and others},
  journal={Artificial Intelligence},
  volume={339},
  pages={104244},
  year={2025},
  publisher={Elsevier}
}

@book{hidalgo2021humans,
  title={How humans judge machines},
  author={Hidalgo, C{\'e}sar A and Orghian, Diana and Canals, Jordi Albo and De Almeida, Filipa and Martin, Natalia},
  year={2021},
  publisher={MIT Press}
}

@article{maslej2025artificial,
  title={Artificial intelligence index report 2025},
  author={Maslej, Nestor and Fattorini, Loredana and Perrault, Raymond and Gil, Yolanda and Parli, Vanessa and Kariuki, Njenga and Capstick, Emily and Reuel, Anka and Brynjolfsson, Erik and Etchemendy, John and others},
  journal={arXiv preprint arXiv:2504.07139},
  year={2025}
}

@article{han2025cooperation,
title = {Cooperation versus social welfare},
journal = {Physics of Life Reviews},
volume = {56},
pages = {33-60},
year = {2026},
issn = {1571-0645},
doi = {https://doi.org/10.1016/j.plrev.2025.11.006},
url = {https://www.sciencedirect.com/science/article/pii/S1571064525001654},
author = {The Anh Han and Zhao Song and Theodor Cimpeanu and Manh Hong Duong and Marcus Krellner and Valerio Capraro and Matjaz Perc}
}

@article{song2026evolution,
  title={Evolution of fairness in hybrid populations with specialised AI agents},
  author={Song, Zhao and Cimpeanu, Theodor and Shen, Chen and Han, The Anh},
  journal={arXiv preprint arXiv:2602.18498},
  year={2026}
}

@article{couto2025collective,
  title={Collective dynamics of strategic classification},
  author={Couto, Marta C and Barsotti, Flavia and Santos, Fernando P},
  journal={arXiv preprint arXiv:2508.09340},
  year={2025}
}

@inproceedings{hardt2016strategic,
  title={Strategic classification},
  author={Hardt, Moritz and Megiddo, Nimrod and Papadimitriou, Christos and Wootters, Mary},
  booktitle={Proceedings of the 2016 ACM Conference on Innovations in Theoretical Computer Science},
  pages={111--122},
  year={2016}
}

@article{mao2017resilient,
  title={Resilient cooperators stabilize long-run cooperation in the finitely repeated prisoner’s dilemma},
  author={Mao, Andrew and Dworkin, Lili and Suri, Siddharth and Watts, Duncan J},
  journal={Nature communications},
  volume={8},
  number={1},
  pages={13800},
  year={2017},
  publisher={Nature Publishing Group UK London}
}





\end{document}